\documentclass[lettersize,journal]{IEEEtran}
\usepackage[T1]{fontenc}
\usepackage{color}
\usepackage{babel}
\usepackage{amsmath,amsfonts}
\usepackage{algorithmic}
\usepackage{algorithm}
\usepackage{array}      
\usepackage{float}
\usepackage{diagbox}
\usepackage{subcaption,caption,epsfig}
\usepackage{makecell} 
\usepackage{pifont} 
\usepackage{multirow}
\usepackage[caption=false,font=normalsize,labelfont=sf,textfont=sf]{subfig}
\usepackage{textcomp}
\usepackage{stfloats}
\usepackage{subcaption} 
\usepackage{url}
\usepackage{svg}
\usepackage{verbatim}
\usepackage{graphicx}
\usepackage{cite}
\usepackage[none]{hyphenat}
\begin{document}

\title{SIaD-Tool: A Comprehensive Frequency-Domain Tool for Small-Signal Stability and Interaction Assessment in Modern Power Systems}

\author{Luis A. Garcia-Reyes,~\IEEEmembership{Graduate Student Member,~IEEE,} Oriol Gomis-Bellmunt,~\IEEEmembership{Fellow Member,~IEEE,} Eduardo Prieto-Araujo,~\IEEEmembership{Senior Member,~IEEE,} Vinícius A. Lacerda, and Marc Cheah-Mañe,~\IEEEmembership{Senior Member,~IEEE}
\vspace{-8mm}
\thanks{L. A. Garcia-Reyes, O. Gomis-Bellmunt, E. Prieto-Araujo, V. A. Lacerda and M. Cheah-Mañe are with Centre d’Innovació Tecnològica en Convertidors Estàtics i Accionaments, Departament d’Enginyeria Elèctrica, Universitat Politècnica de Catalunya (CITCEA-UPC), Barcelona, 08028 ESP (e-mail of corresponding author: luis.reyes@upc.edu).}
}



\maketitle
\vspace{-8mm}
\begin{abstract}
This paper presents SIaD-Tool, an open-source frequency-domain (FD) scanning solution for stability and interaction assessment in modern power systems. The tool enables multi-sequence identification in the \(abc\), \(dq0\), and \(0pn\) frames and supports both series voltage and parallel current perturbation strategies. A novel perturbation scheme allows direct scanning in the target frame, simplifying the analysis of coupling effects and mirrored frequencies. SIaD-Tool is implemented on a multi-platform architecture, including MATLAB/Simulink and Python–PSCAD/EMTDC. Beyond system identification, it integrates automated stability evaluation through four standardized methods: Generalized Nyquist Criterion (GNC), modal impedance analysis, phase margin assessment, and passivity checks. Validation is carried out via extensive case studies involving passive elements, grid-following and grid-forming converters, offshore wind power plants, and the IEEE 9-bus system. Results confirm high accuracy, scalability, and robustness in detecting critical modes, interaction frequencies, oscillatory behavior and stability margins.
\end{abstract}


\section{Introduction}

\IEEEPARstart{M}{odern} power systems are increasingly dominated by power-electronic-based resources (PES), particularly those integrated with renewable energy sources. Technologies such as offshore wind farms, high voltage direct current (HVDC) links, and static synchronous compensators (STATCOMs) enhance grid flexibility and efficiency \cite{paperWIREs}. However, the switching nature of these devices, combined with renewable variability, introduces new challenges—namely, high switching frequencies and complex wide-band controllers—which can trigger subsynchronous and harmonic interactions across multiple system layers \cite{interlinkedInteractions,paperFredeHarmonic}. These interactions span various frequency ranges and may occur between converters and the grid, between adjacent converters, or between converters and passive elements \cite{joaquinpedra}.

Small-signal stability analysis is commonly performed using either state-space eigenvalue analysis or frequency-domain (FD) impedance-based methods. While state-space models provide standardized and accurate results, they require complete mathematical representations of the grid, which becomes impractical for large-scale systems. In contrast, FD methods enable dynamic characterization without explicit models, using electromagnetic transient (EMT) simulations to extract impedance and admittance functions with high accuracy \cite{descriptorstatespacemodeling,passivityandstability,paperWFinteractions}. This approach is particularly advantageous for ``black-box'' systems, where proprietary constraints limit access to internal models \cite{paperBlackBox}, making FD scanning an indispensable tool for system design, validation, and screening.

\vspace{-3.46mm}

\subsection{Literature Review}

FD scanning has become a cornerstone for impedance-based stability analysis in power electronic systems, as it captures wide-band controller dynamics in both AC and DC grids, enabling interaction studies \cite{FDidentificationbook,paperScanner,paperIPST1}. Standardized stability assessment techniques include the Generalized Nyquist Criterion (GNC) \cite{papernyquist}, impedance ratio methods \cite{paperThreePort}, gain and phase margin analysis \cite{papernyquist2,paperBlackBox}, modal impedance analysis \cite{impedance_modal}, and passivity checks \cite{paperJosep}. These approaches provide quantitative metrics to evaluate resonances, grid strength, and multi-frequency interactions in PES-dominated systems \cite{cigre928}.

Despite their maturity, significant challenges persist in consolidating FD methods into flexible, automated, and open-source tools capable of operating seamlessly across multiple platforms. Existing solutions \cite{toolMonash,AIMToolbox,Energinet_IMTB,ZTool_KULeuven} have introduced functionalities such as black-box modeling, multi-terminal scanning, DC-side analysis, and automated impedance measurement; however, they often rely on proprietary environments and fragmented workflows, limiting configurability for comprehensive assessments across reference frames and perturbation strategies. Furthermore, most implementations constrain identification to a single reference frame and fixed perturbation types, reducing adaptability to system-specific dynamics. Combined with platform dependency, licensing restrictions, and the lack of highly detailed standardized documentation, these limitations hinder widespread adoption by academia, transmission system operators (TSOs), and industry stakeholders, ultimately impacting both accuracy and applicability in diverse operational scenarios.

Conventional FD scanning techniques \cite{joaquinpedra,paperIPST2,paperMolinascrosscoupling,paperAgustiCrosscouplping} reveal coupling between sequences, requiring complex mapping schemes to identify impedance and admittance elements. Converter and system asymmetries further exacerbate these challenges, especially given the lack of standardized methods to capture coupling effects. Perturbation strategies such as parallel current and series voltage injections \cite{paperWFinteractions,paperBlackBox} achieve high accuracy up to 100~Hz for single operating points in the synchronous \( dq0 \) frame. Sinusoidal single-tone signals \cite{dist_signals_chemistry} maximize signal-to-noise ratio compared to pseudo-random \cite{paperseudorandom} and multi-tone excitations \cite{multisine}, while emerging parametric identification methods \cite{paperVerena} aim to accelerate scanning over broader frequency ranges but remain limited by converter-centric identification.

\vspace{-4mm}
\subsection{Contributions}

To overcome the limitations of existing impedance-based FD tools and methodologies, the proposed \textit{Stability and Interactions assessment in the frequency-Domain} tool (SIaD-Tool) introduces an automated, flexible, open-source framework for impedance and admittance scanning across multiple reference frames (\( abc \), \( 0pn \), and \( dq0 \)). It leverages a novel steady-state and injection scheme that simplifies mirror-frequency complexities and supports seamless integration with platforms such as PSCAD, Python, and MATLAB/Simulink through openly documented methodology and source code. Unlike previous solutions, SIaD-Tool is not restricted to a single simulation environment and offers full configurability in perturbation strategies and excitation signals, enabling tailored scanning for diverse system dynamics. Its modular design provides advanced capabilities:
\begin{enumerate}
    \item Identification of interaction frequencies and stability margins across wide frequency ranges, covering subsynchronous and harmonic regimes.
    \item Prediction of oscillatory and damping behavior.
    \item Multi-operating point and control tuning analysis.
    \item Assessment of interactions between active and passive components in AC/DC systems.
\end{enumerate}

A comparative summary of SIaD-Tool capabilities against other tools is presented in Table~\ref{tab:tools}. The tool is freely accessible in the repository \cite{SIaD_tool}, promoting transparency, reproducibility, and broader adoption of impedance-based stability studies in power electronics-dominated systems.

\begin{table}[t!]
\centering
\caption{Comparison of available tools for impedance-based stability analysis in modern power systems.}
\label{tab:tools}

\renewcommand\arraystretch{1.15}

\begin{tabular}{| >{\raggedright}m{2.6cm} | c | c | c | c |}
\hline
\diagbox[width=3cm]{\textbf{Capabilities}}{\textbf{Tool}}
& \makecell{\textbf{AIStability}\\\cite{AIMToolbox}}
& \makecell{\textbf{IMTB}\\\cite{Energinet_IMTB}}
& \makecell{\textbf{Z-tool}\\\cite{ZTool_KULeuven}}
& \makecell{\textbf{SIaD-Tool}\\\cite{SIaD_tool}} \\ \hline

Identification in multiple sequences     & \ding{55} & \ding{55} & \ding{55} & \ding{51} \\ \hline
Different disturbance strategies         & \ding{55} & \ding{51} & \ding{55} & \ding{51} \\ \hline
Different perturbation signals           & \ding{55} & \ding{55} & \ding{51} & \ding{51} \\ \hline
DC-side identification  & \ding{55} & \ding{51} & \ding{51} & \ding{51} \\ \hline
GNC stability analysis                   & \ding{51} & \ding{55} & \ding{51} & \ding{51} \\ \hline
Stability analysis with gain and phase margin & \ding{55} & \ding{55} & \ding{55} & \ding{51} \\ \hline
Modal impedance analysis                 & \ding{55} & \ding{55} & \ding{51} & \ding{51} \\ \hline
Passivity assessment                     & \ding{55} & \ding{55} & \ding{51} & \ding{51} \\ \hline
Stability margin index                   & \ding{55} & \ding{55} & \ding{55} & \ding{51} \\ \hline
Multi-platform tool                         & \ding{55} & \ding{55} & \ding{55} & \ding{51} \\ \hline
Open-source tool                         & \ding{55} & \ding{51} & \ding{51} & \ding{51} \\ \hline
\end{tabular}
\vspace{-4mm}
\end{table}
\vspace{-3mm}
\subsection{Paper Structure}

The paper is organized as follows: Section~II details the methodology for FD identification, including perturbation strategies and multi-sequence scanning. Section III validates the FD scanning tool in linear power system elements. Section~IV describes the stability and interaction assessment methodologies, covering GNC, modal impedance, phase margin, and passivity analysis. Section~V validates the tool through multiple case studies, including converter models, offshore wind systems, and the IEEE 9-bus system. Finally, Section~VI concludes the paper and outlines future work.

\vspace{-3mm}

\section{Methodology for FD Identification}\label{sec:mathematical}

The FD scanning technique characterizes system impedance or admittance through EMT simulations by injecting small-amplitude voltage or current perturbations at specific or multiple frequencies at the Point of Scanning (PoS). These perturbations are implemented as series voltage sources or parallel current sources, while preserving equilibrium with the steady-state source operating at the nominal frequency and maintaining the system initial conditions \cite{impedancebook}.

\begin{figure}[t!]
 \centering
 \includegraphics[width=0.7\columnwidth]{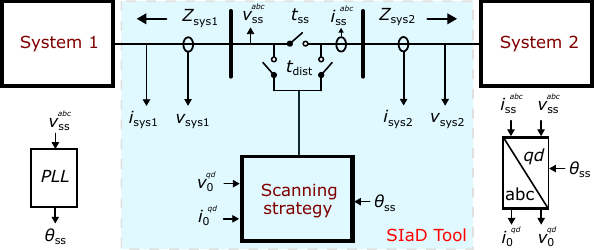}
 \vspace{-1mm}
 \caption{SIaD-Tool architecture.}
 \label{fig:siad}
 \vspace{-4mm}
\end{figure}

Unlike other tools \cite{AIMToolbox,ZTool_KULeuven}, SIaD-Tool decouples and re-couples the system using ideal sources that replicate pre-decoupling operating conditions. Voltage magnitude and angle are extracted from a prior measurement stage to ensure consistency. Fig.~\ref{fig:siad} illustrates the operational scheme, which comprises the following stages:

\paragraph{Steady-State Definition}
In this stage, voltage, current, and phase angle at the decoupling point are measured (\( t_{ss} \) ON, \( t_{dist} \) OFF). These quantities are expressed in the \( dq \) frame for balanced systems or in the zero-positive-negative (\( 0pn \)) frame for unbalanced systems. Alternatively, this step can be skipped if steady-state data are provided from power flow results. Since the converter only observes conditions at the PoS, the ideal source can reproduce the same power flow profile without introducing deviations \cite{dqsspoint}.

\paragraph{Side-to-Side Identification}
The system is decoupled at the PoS, isolating one side from the other (\( t_{ss} \) OFF, \( t_{dist} \) ON). The FD scanner is then connected to perform the identification process using steady-state data from the previous stage to reconstruct operating conditions. The desired perturbation strategy and signal type are applied in this stage.

\begin{figure}[t!]
    \centering
    \begin{subfigure}[t]{0.5\columnwidth}
        \centering
        \includegraphics[width=\textwidth]{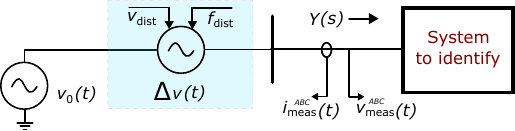}
        \vspace{-4mm}
        \caption{Series voltage.}
        \label{fig:injection_voltage}
    \end{subfigure}
    \vspace{0.5em}
    \begin{subfigure}[t]{0.53\columnwidth}
        \centering
        \includegraphics[width=\textwidth]{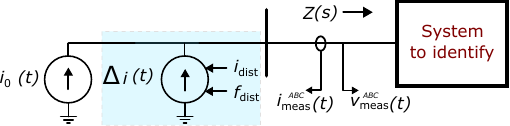}
        \vspace{-4mm}
        \caption{Parallel current.}
        \label{fig:injection_current}
    \end{subfigure}
    \vspace{-2mm}
    \caption{Time-domain scanning strategies.}
    \label{fig:injection}
    \vspace{-4mm}
\end{figure}

Fig.~\ref{fig:injection} shows the two time-domain (TD) injection strategies implemented in SIaD-Tool\cite{paperWFinteractions,paperScanner,impedancebook}. Voltage and current waveforms at the PoS are recorded in the TD prior to decoupling. These measurements are essential for computing deviations from steady-state and determining the small-signal quantities required for identification. 

Perturbation signals are applied over a specific time interval or throughout the simulation and measured within a steady-state window, which defines signal length and spectral resolution. TD signals are converted to the FD via the Fast Fourier Transform (FFT), and impedance or admittance values are iteratively calculated for each perturbation type and strategy using algebraic FD operations. For identification on the DC-side a special procedure is provided in \cite{SIaD_tool}.

\vspace{-3mm}
\subsection{Perturbation Signals}
SIaD-Tool integrates several perturbation signals to identify FD matrices, each offering specific advantages and limitations:

\subsubsection{Single-Tone Signal}

This method uses a sinusoidal waveform with fixed amplitude and frequency, as defined in \eqref{eq:singletone}, and requires $M$ separate simulations to analyze $M$ distinct frequencies. Its main advantage lies in its high signal-to-noise ratio, which enables accurate extraction of frequency components with minimal spectral leakage.
\begin{subequations}\label{eq:singletone}
\begin{align}
    \Delta v(t) &= V_d \cos(2\pi f_d t), \label{eq:singletone_v} \\
    \Delta i(t) &= I_d \cos(2\pi f_d t). \label{eq:singletone_i}
\end{align}
\end{subequations}

Here, $\Delta v$ and $\Delta i$ denote voltage and current perturbations, with peak amplitudes $V_d$ and $I_d$ (typically 1--3\% of nominal values \cite{paperBlackBox}) to ensure small-signal operation. The disturbance frequency is $f_d$, with angular frequency $\omega_d = 2\pi f_d$.

\subsubsection{Multi-Tone Signal}

This approach superposes multiple cosine components, enabling simultaneous injection of several frequencies and reducing the number of required simulations. Its main drawback is that nonlinearities in power electronic systems may introduce distortions, compromising frequency estimation accuracy \cite{dist_signals_chemistry}. The implemented signal follows the algorithm in \cite{holmes_multisine} and is expressed as:
\begin{subequations}\label{eq:multitone}
\begin{align}
    \Delta v(t) &= \sum_{k=1}^{N} V_{d_k}\cos(2\pi f_k t + \varphi_k), \label{eq:multitone_v} \\
    \Delta i(t) &= \sum_{k=1}^{N} I_{d_k}\cos(2\pi f_k t + \varphi_k). \label{eq:multitone_i}
\end{align}
\end{subequations}

Here, \( N \) is the number of components, while \( f_k \) and \( \varphi_k \) denote the frequency and phase of the \( k \)-th tone. This formulation offers high flexibility for multi-frequency excitation design. A critical constraint is minimizing the crest factor to prevent nonlinear behavior; the algorithm in \cite{holmes_multisine} applies phase optimization for this purpose. Additional design guidelines are provided in \cite{multisine}.

\subsubsection{Pseudo-Random Binary Sequence (PRBS)}

A PRBS is a deterministic binary signal that mimics the statistical properties of white noise over a finite time window. It is generated using a linear feedback shift register (LFSR), governed by the following difference equations:
\begin{subequations}\label{eq:prbs}
\begin{align}
    \Delta v(t) &= \mathrm{rem}\left( a_1 v(t-1) + \dots + a_n v(t-n),\ 2 \right), \label{eq:prbs_v} \\
    \Delta i(t) &= \mathrm{rem}\left( a_1 i(t-1) + \dots + a_n i(t-n),\ 2 \right), \label{eq:prbs_i}
\end{align}
\end{subequations}

Despite its efficiency, PRBS signals may exhibit low signal-to-noise ratios and uneven frequency excitation, which can affect identification precision. In such cases, filtering may be required to accurately extract the FD response \cite{paperseudorandom}.


\vspace{-5mm}
\subsection{$abc$ Sequence Scanning}

This subsection describes the phase sequence identification procedure based on the scanning methodology previously defined for both voltage and current strategies \cite{cigre928}.

\subsubsection{Voltage Disturbance Method}

The admittance response in the $abc$ phase domain is determined using the scheme illustrated in Fig.~\ref{fig:injection_voltage}. Prior to system decoupling, voltage and current waveforms at the PoS are recorded within the desired time window and stored as \( \mathbf{v}^{abc}_{ss}(t) \) and \( \mathbf{i}^{abc}_{ss}(t) \), respectively. These signals represent the steady-state operating condition and serve as references for computing small-signal deviations during the identification process.

Once the system is decoupled, the fundamental source is redefined using an ideal voltage source that replicates the pre-disturbance steady-state condition. This source is expressed in its generalized form as:
\begin{equation}\label{eq:fundamentalsourceV}
\mathbf{v}_{0}(t) =
\left\{
\begin{array}{l}
\mathrm{V}_{\text{peak}} \cos \left(\omega_0 t + \phi_a\right) \\
\mathrm{V}_{\text{peak}} \cos \left(\omega_0 t - \frac{2\pi}{3} + \phi_b\right) \\
\mathrm{V}_{\text{peak}} \cos \left(\omega_0 t + \frac{2\pi}{3} + \phi_c \right)
\end{array}
\right. ,
\end{equation}
where \( \omega_0 = 2\pi f_0 \), with \( f_0 \) denoting the fundamental frequency. Here, \( \mathrm{V}_{\text{peak}} \) is the peak voltage of the system, and \( \phi_a, \phi_b, \phi_c \) are the initial phase angles of each phase. These values must be measured or specified prior to decoupling.

The single- or multi-tone disturbance method applies perturbations independently to the $a$, $b$, and $c$ phases. The $abc$ perturbation source with the measured angle $\theta_m$ is defined as:
\begin{equation}\label{eq:disturbancesourceVABC}
\Delta \mathbf{v}(t) = \boldsymbol{\eta}_{x, \mathrm{dist}} \, \Delta v(t),
\end{equation}
where \( \boldsymbol{\eta}_{x, \mathrm{dist}} \) is the corresponding vector to the excited phase:
\begin{equation}\label{eq:nx}
\begin{array}{ccc}
\boldsymbol{\eta}_{a, \mathrm{dist}}=
\begin{bmatrix}
    1 \\
    0 \\
    0
\end{bmatrix}, &
\boldsymbol{\eta}_{b, \mathrm{dist}}=
\begin{bmatrix}
    0 \\
    1 \\
    0
\end{bmatrix}, &
\boldsymbol{\eta}_{c, \mathrm{dist}}=
\begin{bmatrix}
    0 \\
    0 \\
    1
\end{bmatrix}
\end{array}
\end{equation}

After applying the disturbance and reaching steady-state, a time window is applied to the measured signals:
\begin{subequations}\label{eq:windowedsignalsABC}
\begin{align}
    \mathbf{v}_{abc}(t) &= w(t) \odot \mathbf{v}^{abc}_{\text{meas}}(t), \label{eq:windowedsignals1} \\
    \mathbf{i}_{abc}(t) &= w(t) \odot \mathbf{i}^{abc}_{\text{meas}}(t), \label{eq:windowedsignals2}
\end{align}
\end{subequations}
where \( \odot \) denotes the element-wise product, and the window function \( w(t) \) is defined as:
\begin{equation}\label{eq:window}
w(t) =
\begin{cases} 
1, & t_{\text{start}} \leq t \leq t_{\text{end}}, \\ 
0, & \text{otherwise}.
\end{cases}
\end{equation}

The time window determines the frequency resolution:
\begin{equation}\label{eq:fres}
f_{\text{res}} = {1}/{\left(t_{\text{end}} - t_{\text{start}}\right)},
\end{equation}
where \( t_{\text{start}} \) and \( t_{\text{end}} \) define the window limits. The FFT is then applied to the signals, subtracting the steady-state contribution:
\vspace{-4mm}
\begin{subequations}\label{eq:FFTsignals}
\begin{align}
    \Delta \mathbf{V}_{abc}(j \omega) &= \text{FFT} \{ \mathbf{v}_{abc}(t) - \mathbf{v}^{abc}_{ss}(t) \}, \label{eq:fftsignals1} \\
    \Delta \mathbf{I}_{abc}(j \omega) &= \text{FFT} \{ \mathbf{i}_{abc}(t) - \mathbf{i}^{abc}_{ss}(t) \}. \label{eq:fftsignals2}
\end{align}
\end{subequations}

For single-tone perturbations, each admittance matrix element is extracted at the disturbance frequency \( f_d \), using:
\begin{equation}\label{eq:freqindex}
\Omega_d = {f_d}/{f_{\text{res}}},
\end{equation}
where \( f_d \) must be a multiple of \( f_{\text{res}} \), ensuring \( \Omega_d \in \mathbb{Z}^+ \). Additionally, \( f_d \) must satisfy the Nyquist criterion:
\begin{equation}\label{eq:frule}
f_d \leq {f_{\text{sampling}}}/{2},
\end{equation}
with \( f_{\text{sampling}} = 1/\Delta t \), where \( \Delta t \) is the EMT simulation step time. For industrial applications, it is recommended to limit \( f_d \leq f_{\text{sampling}}/4 \) to maintain an adequate signal-to-noise ratio.

The admittance matrix \( \mathbf{Y}_{abc}(j \omega_d) \) is constructed as:
\begin{equation}\label{eq:YmatrixABC_Vdist}
Y_{xy}(\Omega_d) = {\Delta I_x(\Omega_d)}/{\Delta V_y(\Omega_d)},
\end{equation}
where \( x, y \in \{a, b, c\} \). Depending on the injected phase:
\begin{itemize}
    \item \( a \)-injection: compute elements \( y = a \) and \( x \in \{a,b,c\} \).
    \item \( b \)-injection: compute elements \( y = b \) and \( x \in \{a,b,c\} \).
    \item \( c \)-injection: compute elements \( y = c \) and \( x \in \{a,b,c\} \).
\end{itemize}

The same principle applies to multi-tone and PRBS signals, provided the frequency band satisfies \eqref{eq:frule}. In these cases, TD measurements from \eqref{eq:windowedsignalsABC} are processed via \eqref{eq:FFTsignals} and \eqref{eq:YmatrixABC_Vdist}. Unlike the single-tone approach, computation occurs once per phase injection, and the admittance matrix is assembled by stacking the resulting vectors into a three-dimensional structure.

\subsubsection{Current Disturbance Method}

The \( abc \) impedance matrix is identified using the scheme illustrated in Fig.~\ref{fig:injection_current}, with the fundamental current source defined as:
\begin{equation}\label{eq:fundamentalsourceI}
\mathbf{i}_{0}(t) =
\left\{
\begin{array}{l}
\mathrm{I}_{\text{peak}} \cos \left(\omega_0 t + \phi_a\right) \\
\mathrm{I}_{\text{peak}} \cos \left(\omega_0 t - \frac{2\pi}{3} + \phi_b\right) \\
\mathrm{I}_{\text{peak}} \cos \left(\omega_0 t + \frac{2\pi}{3} + \phi_c\right)
\end{array}
\right. ,
\end{equation}
where \( \mathrm{I}_{\text{peak}} \) is the peak current at the PoS, and \( \phi_a, \phi_b, \phi_c \) are the initial phase angles. The disturbance source is given by:
\begin{equation}\label{eq:disturbancesourceIABC}
\Delta \mathbf{i}(t) = \boldsymbol{\eta}_{x, \mathrm{dist}} \, \Delta i(t),
\end{equation}
where \( \boldsymbol{\eta}_{x, \mathrm{dist}} \) is from \eqref{eq:nx}, and \( \Delta i(t) \) is the perturbation signal.

Following the procedure described in \cite{paperImethod,cigre928}, voltage and current measurements are extracted within the defined time window for each independent phase injection, this is: for phase \( a \)-injection, \( i_{xa} \) and \( v_{xa} \) are recorded; for phase \( b \)-injection, \( i_{xb} \) and \( v_{xb} \); and for phase \( c \)-injection, \( i_{xc} \) and \( v_{xc} \), where \( x \in \{a, b, c\} \) denotes the measured phase.

For single-tone perturbations, the small-signal voltage and current matrices in the FD are constructed from the measurements processed by \eqref{eq:FFTsignals} and indexed using \eqref{eq:freqindex}:
{\small
\begin{subequations}\label{eq:VImeas_Imethod}
\begin{align}
\Delta \mathbf{V}_{abc}(\Omega_d) &=
\begin{pmatrix}
\Delta V_{aa}(\Omega_d) & \Delta V_{ab}(\Omega_d) & \Delta V_{ac}(\Omega_d) \\
\Delta V_{ba}(\Omega_d) & \Delta V_{bb}(\Omega_d) & \Delta V_{bc}(\Omega_d) \\
\Delta V_{ca}(\Omega_d) & \Delta V_{cb}(\Omega_d) & \Delta V_{cc}(\Omega_d)
\end{pmatrix}, \label{eq:Vmeas_Imethod} \\
\Delta \mathbf{I}_{abc}(\Omega_d) &=
\begin{pmatrix}
\Delta I_{aa}(\Omega_d) & \Delta I_{ab}(\Omega_d) & \Delta I_{ac}(\Omega_d) \\
\Delta I_{ba}(\Omega_d) & \Delta I_{bb}(\Omega_d) & \Delta I_{bc}(\Omega_d) \\
\Delta I_{ca}(\Omega_d) & \Delta I_{cb}(\Omega_d) & \Delta I_{cc}(\Omega_d)
\end{pmatrix}. \label{eq:Imeas_Imethod}
\end{align}
\vspace{-2mm}
\end{subequations}
}
The impedance matrix is then computed as:
\begin{equation}\label{eq:ZmatrixABC_Idist}
\mathbf{Z}_{abc}(\Omega_d) = \Delta \mathbf{V}_{abc}(\Omega_d) \left[ \Delta \mathbf{I}_{abc}(\Omega_d) \right]^{-1}.
\end{equation}

For multi-tone and PRBS, FD measurements from \eqref{eq:windowedsignalsABC} are processed using \eqref{eq:FFTsignals} and \eqref{eq:ZmatrixABC_Idist}. In these cases, voltage and current matrices are assembled for each frequency sample to compute the impedance matrix on a sample-by-sample basis.
\vspace{-3mm}
\subsection{$0pn$ Sequence Scanning}

This subsection describes the $0pn$ sequence identification procedure based on the scanning methodology previously defined for both disturbance strategies.

\subsubsection{Voltage Disturbance Method}

The same fundamental source defined in \eqref{eq:fundamentalsourceV} is used when the system is balanced. Otherwise, an unbalanced source must be constructed as:
\begin{equation}\label{eq:fundamentalsourceV_0pn}
\mathbf{v}_0(t) =  
\mathbf{T}^{-1}_{0pn} 
\begin{bmatrix}
 \mathrm{V}_0 \ e^{j \theta_0} \\
 \mathrm{V}_p \ e^{j \theta_p} \\
 \mathrm{V}_n \ e^{j \theta_n}
\end{bmatrix} e^{j \omega t},
\end{equation}
where \( V_0 \), \( V_p \), and \( V_n \) are the phasor magnitudes, and \( \theta_0 \), \( \theta_p \), and \( \theta_n \) are the phase angles of the zero, positive, and negative sequence components. The inverse Fortescue transform is:
\begin{equation}
    \mathbf{T}^{-1}_{0pn} = \begin{bmatrix}
1 & 1 & 1 \\
1 & e^{j2\pi/3} & e^{-j2\pi/3} \\
1 & e^{-j2\pi/3} & e^{j2\pi/3}
\end{bmatrix}.
\end{equation}

The disturbance source is:
\begin{equation}\label{eq:disturbancesourceVpn0}
\Delta \mathbf{v}(t) =  
\mathbf{T}^{-1}_{0pn} \left[ \boldsymbol{\eta}_{x, \mathrm{dist}} \ \Delta v(t) \right],
\end{equation}
where \( \boldsymbol{\eta}_{x, \mathrm{dist}} \) is the vector from \eqref{eq:nx} in the \( 0pn \) domain.

Prior to decoupling, steady-state voltage and current waveforms are recorded and stored as \( \mathbf{v}^{0pn}_{ss}(t) \) and \( \mathbf{i}^{0pn}_{ss}(t) \). Unlike the $abc$-frame scanning, these signals are directly obtained in the $0pn$ sequence as complex vectors.

After applying the disturbance and reaching steady-state, voltage and current measurements are taken at the PoS and transformed to the $0pn$ domain:
\begin{subequations}\label{eq:0pnmeas_Vmethod}
\begin{align}
    \mathbf{v}^{0pn}_{\text{meas}}(t) &= \mathbf{T}_{0pn} \, \mathbf{v}^{abc}_{\text{meas}}(t), \label{eq:Vmeaspn0} \\
    \mathbf{i}^{0pn}_{\text{meas}}(t) &= \mathbf{T}_{0pn} \, \mathbf{i}^{abc}_{\text{meas}}(t). \label{eq:Imeaspn0}
\end{align}
\end{subequations}

Here the time window is applied:
\begin{subequations}\label{eq:0pn_windowedsignals}
\begin{align}
    \mathbf{v}_{0pn}(t) &= w(t) \odot \mathbf{v}^{0pn}_{\text{meas}}(t), \label{eq:opn_windowedsignals1} \\
    \mathbf{i}_{0pn}(t) &= w(t) \odot \mathbf{i}^{0pn}_{\text{meas}}(t), \label{eq:opn_windowedsignals2}
\end{align}
\end{subequations}
and the FD signals are computed as:
\begin{subequations}\label{eq:0pnspectra_Vmethod}
\begin{align}
    \Delta \mathbf{V}_{0pn}(j \omega) &= \text{FFT}[ \mathbf{v}_{0pn}(t) - \mathbf{v}^{0pn}_{ss}(t)], \label{eq:fftsiganls0pn1} \\
    \Delta \mathbf{I}_{0pn}(j \omega) &= \text{FFT}[ \mathbf{i}_{0pn}(t) - \mathbf{i}^{0pn}_{ss}(t)]. \label{eq:fftsiganls0pn2}
\end{align}
\end{subequations}

\begin{table}[t!]
\scriptsize
\setlength{\tabcolsep}{2pt}
\centering
\caption{Conditions for $pn$ sequence perturbation.}
\label{tab:mirror}
\resizebox{\columnwidth}{!}{
\begin{tabular}{|c|c|c|c|}
\hline
 & \multicolumn{2}{c|}{\textbf{Injection in p-frame}} & \textbf{Injection in n-frame} \\ \hline
 & $0 < f_d < 2f_0$ & $f_d \geq 2f_0$ & $f_d > 0$ \\ \hline
$Y_{pp}(\Omega_d)$ & \multicolumn{2}{c|}{$\frac{I_p (\Omega_{d})}{V_p (\Omega_{d})}$} & -- \\ \hline
$Y_{np}(\Omega_{d})$ & $\frac{I_n (\Omega_{d-p})}{V_p (\Omega_{d})}$ & $\frac{I_n (-\Omega_{d-p})}{V_p (\Omega_{d})}$ & -- \\ \hline
$Y_{pn}(\Omega_{d})$ & -- & -- & $\frac{I_p (\Omega_{d-n})}{V_n (\Omega_{d})}$ \\ \hline
$Y_{nn}(\Omega_{d})$ & -- & -- & $\frac{I_n (\Omega_{d})}{V_n (\Omega_{d})}$ \\ \hline
\end{tabular}
} \vspace{-3mm}
\end{table}

Unlike conventional methods, SIaD-Tool generates the disturbance source directly in the \(0pn\)-frame, allowing the admittance matrix to be obtained in the same frame for passive elements. For PES, the mirroring scheme in Table~\ref{tab:mirror} is applied, eliminating mirrored-frequency artifacts that arise when injecting in the \(abc\)-frame and measuring in \(0pn\) \cite{paperMolinascrosscoupling}. The frequency index in \eqref{eq:freqindex} is then adapted as:
\begin{equation}\label{eq:freqindex_1}
\Omega_{d-p} = \frac{2f_0-f_d}{f_{\text{res}}};  \quad \Omega_{d-n} = \frac{f_d+2f_0}{f_{\text{res}}}.
\end{equation}
Notice that the remaining terms are obtained using a procedure similar to that described in \eqref{eq:YmatrixABC_Vdist} using the $0pn$ sequence. 

\subsubsection{Current Disturbance Method}

For balanced systems, the fundamental source is defined in \eqref{eq:fundamentalsourceI}; for unbalanced systems:
\begin{equation}\label{eq:fundamentalsourceI_0pn}
\mathbf{i}_0(t) =  
\mathbf{T}^{-1}_{0pn} 
\begin{bmatrix}
 \mathrm{I}_0 \ e^{j \theta_0} \\
 \mathrm{I}_p \ e^{j \theta_p} \\
 \mathrm{I}_n \ e^{j \theta_n}
\end{bmatrix} e^{j \omega t},
\end{equation}
where \( \mathrm{I}_0 \), \( \mathrm{I}_p \), and \( \mathrm{I}_n \) are the phasor magnitudes of the $0$, $p$, and $n$ sequence components. The disturbance source is defined as:
\begin{equation}\label{eq:disturbancesourceIpn0}
\Delta \mathbf{i}(t) = \mathbf{T}^{-1}_{0pn} \left[ \boldsymbol{\eta}_{x, \mathrm{dist}} \ \Delta i(t) \right].
\end{equation}

The voltage and current matrices \( \Delta \mathbf{V}_{0pn}(\Omega_d) \) and \( \Delta \mathbf{I}_{0pn}(\Omega_d) \) are constructed using the same procedure described in \eqref{eq:0pnmeas_Vmethod}–\eqref{eq:freqindex_1}, extracting the spectra for each independent injection and removing the steady-state contribution. The impedance matrix is then computed:
\begin{equation}\label{eq:Zmatrix0pn_Idist}
\mathbf{Z}_{0pn}(\Omega_d) = \Delta \mathbf{V}_{0pn}(\Omega_d) \left[ \Delta \mathbf{I}_{0pn}(\Omega_d) \right]^{-1}.
\end{equation}

\vspace{-4mm}

\subsection{$dq0$ Sequence Scanning}

For scanning in the \( dq0 \)-frame, the fundamental sources are synthesized directly in the \( dq0 \)-frame. As in previous methods, steady-state voltage and current at the PoS, denoted \( \mathbf{v}^{dq0}_{ss}(t) \) and \( \mathbf{i}^{dq0}_{ss}(t) \), are recorded prior to system decoupling. The reference angle required for proper \( dq0 \)-frame alignment is:
\begin{equation}\label{eq:thetaref}
\theta(t) = \omega t + \theta_0=\int_{0}^{t} 2\pi f_0 \, d\tau + \theta_0,
\end{equation}
where \( f_0 \) is the fundamental frequency, \( \tau \) is the integration variable, and \( \theta_0 \) is the measured steady-state angle (Fig. \ref{fig:siad}).

\subsubsection{Voltage Disturbance Method}

The fundamental voltage source in the \( dq0 \) frame is defined as:
\begin{equation}\label{eq:fundamentalsourceV_dq0}
\mathbf{v}_0(t) =  
\mathbf{T}^{-1}_{dq0}(\theta) 
\begin{bmatrix}
 \mathrm{V}_d  \\
 \mathrm{V}_q  \\
 \mathrm{V}_0 
\end{bmatrix}, \quad \text{with} \quad \theta(t) = \omega t + \theta_0,
\end{equation}
and the corresponding disturbance source is given by:
\begin{equation}\label{eq:disturbancesourceVdq0}
\Delta \mathbf{v}(t) =  
\mathbf{T}^{-1}_{dq0} (\theta)
\begin{bmatrix}
\boldsymbol{\eta}_{x, \mathrm{dist}} \ \Delta v(t)
\end{bmatrix},
\end{equation}
where \( \mathbf{T}^{-1}_{dq0}(\theta) \) is the inverse Park transformation matrix:
\begin{equation}\label{eq:Tdq0}
\mathbf{T}^{-1}_{dq0} (\theta) = 
\begin{bmatrix}
\cos\left(\theta\right) & \sin\left(\theta\right) & 1 \\
\cos\left(\theta - \frac{2\pi}{3}\right) & \sin\left(\theta - \frac{2\pi}{3}\right) & 1 \\
\cos\left(\theta + \frac{2\pi}{3}\right) & \sin\left(\theta + \frac{2\pi}{3}\right) & 1
\end{bmatrix}.
\end{equation}

After applying the disturbance and reaching steady-state, voltage and current measurements are taken in the \( dq0 \)-frame using \eqref{eq:Tdq0}, following a procedure analogous to that described in \eqref{eq:0pnmeas_Vmethod}. The TD signals are then windowed and transformed using the FFT, as described in \eqref{eq:0pn_windowedsignals}–\eqref{eq:0pnspectra_Vmethod}, considering the steady-state contributions in the \( dq0 \)-frame, denoted by \( \mathbf{v}^{dq0}_{ss}(t) \) and \( \mathbf{i}^{dq0}_{ss}(t) \). Finally, the admittance matrix \( \mathbf{Y}_{dq0}(j \omega_d) \) is computed using a procedure analogous to \eqref{eq:YmatrixABC_Vdist}.

\subsubsection{Current Disturbance Method}
The fundamental and the disturbance source are given by:
\begin{equation}\label{eq:fundamentalsourceI_dq0}
\mathbf{i}_0(t) =  
\mathbf{T}^{-1}_{dq0}(\theta) 
\begin{bmatrix}
 \mathrm{I}_d  \\
 \mathrm{I}_q  \\
 \mathrm{I}_0 
\end{bmatrix}, \quad \text{with} \quad \theta(t) = \omega t + \theta_0,
\end{equation}
\begin{equation}\label{eq:disturbancesourceIdq0}
\Delta \mathbf{i}(t) =  
\mathbf{T}^{-1}_{dq0} (\theta)
\begin{bmatrix}
\boldsymbol{\eta}_{x, \mathrm{dist}} \ \Delta i(t)
\end{bmatrix}.
\end{equation}

The voltage and current matrices \( \Delta \mathbf{V}_{dq0}(\varOmega_d) \) and \( \Delta \mathbf{I}_{dq0}(\varOmega_d) \) are assembled as in \eqref{eq:VImeas_Imethod}, and the impedance matrix is computed following the procedure in \eqref{eq:ZmatrixABC_Idist}.

\vspace{-5mm}

\section{Frequency-Domain Scanning Validation}\label{sec:validation}

\begin{figure}[t!]
    \centering
    \begin{subfigure}[b]{0.4\columnwidth}
        \centering
\includegraphics[width=\textwidth]{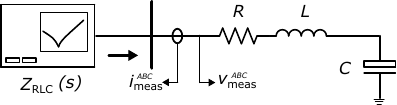}
        \vspace{-5mm}
        \caption{RLC load.}
        \label{fig:RLCloadscheme}
    \end{subfigure}
    \hfill
    \begin{subfigure}[b]{0.6\columnwidth}
        \centering
\includegraphics[width=\textwidth]{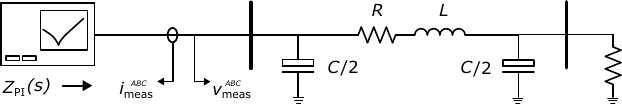}
        \vspace{-5mm}
        \caption{PI section.}
        \label{fig:PIsectionscheme}
    \end{subfigure}
    \vspace{0.5em}
    \begin{subfigure}[b]{0.6\columnwidth}
        \centering
\includegraphics[width=\textwidth]{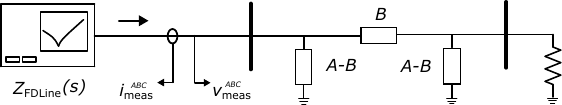}
        \vspace{-5mm}
        \caption{Frequency-dependent line.}
        \label{fig:FDlinescheme}
    \end{subfigure}
    \vspace{-2mm}
    \caption{General scanning scheme for passive elements.}
    \label{fig:passiveschemes}
    \vspace{-4mm}
\end{figure}

The proposed multi-sequence, multi-strategy scanning methodology was validated through three MATLAB/Simulink case studies, with a Python/PSCAD implementation available in the public repository \cite{SIaD_tool}. Validation scenarios include: A) identification in the \( abc \) frame of a series \( RLC \) load using various perturbation signals; B) identification in the \( dq0 \) frame of a PI-equivalent cable model; and C) identification in the \( 0pn \) frame of a frequency-dependent transmission line. The general simulation scheme is shown in Fig.~\ref{fig:passiveschemes}.

\vspace{-4mm}

\subsection{RLC Scanning}

\begin{figure}[t!]
    \centering
    \begin{subfigure}[b]{0.35\columnwidth}
        \centering
        \includegraphics[width=\textwidth]{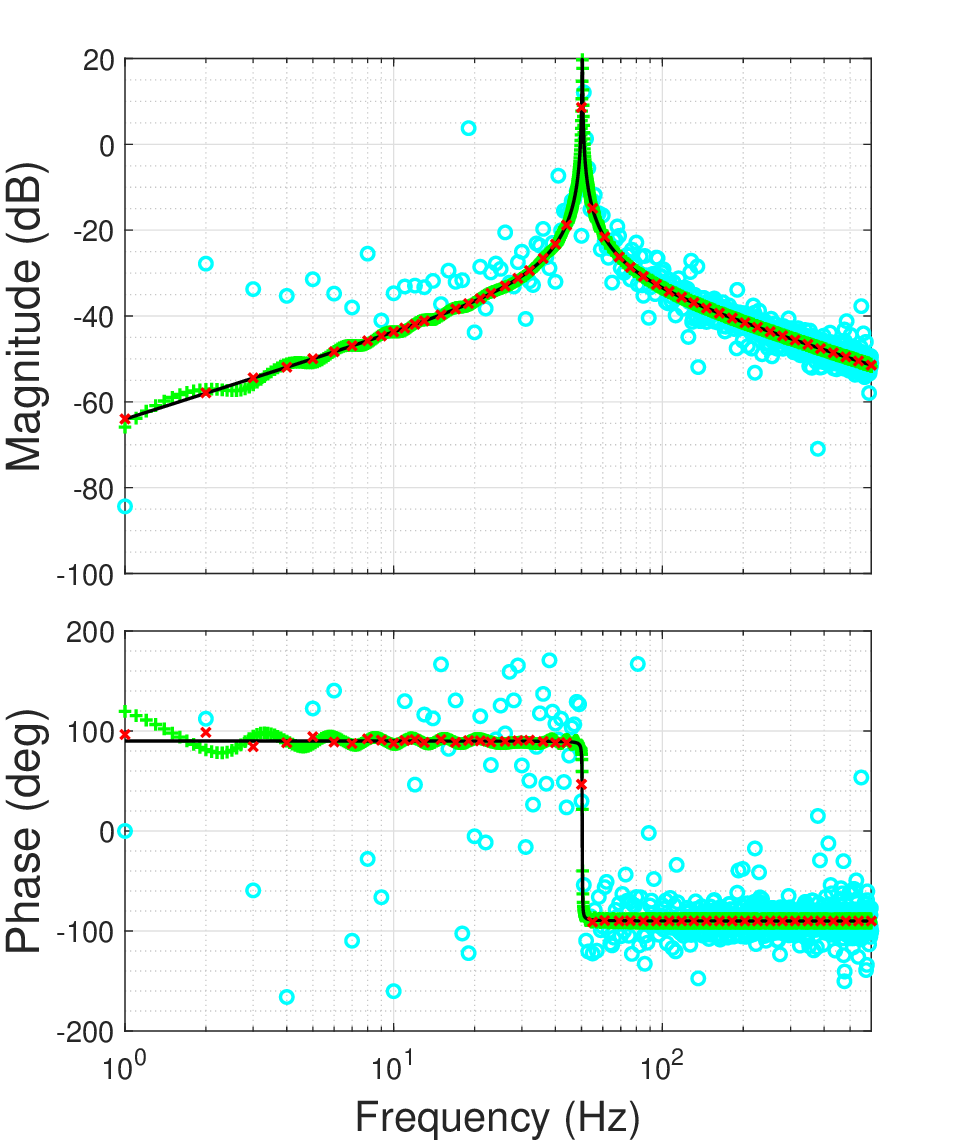}
        \vspace{-5mm}
        \caption{\( Y_{abc} \) with voltage perturbation.}
        \label{fig:RLCa}
    \end{subfigure}
    \hspace{-1mm}
    \begin{subfigure}[b]{0.35\columnwidth}
        \centering
        \includegraphics[width=\textwidth]{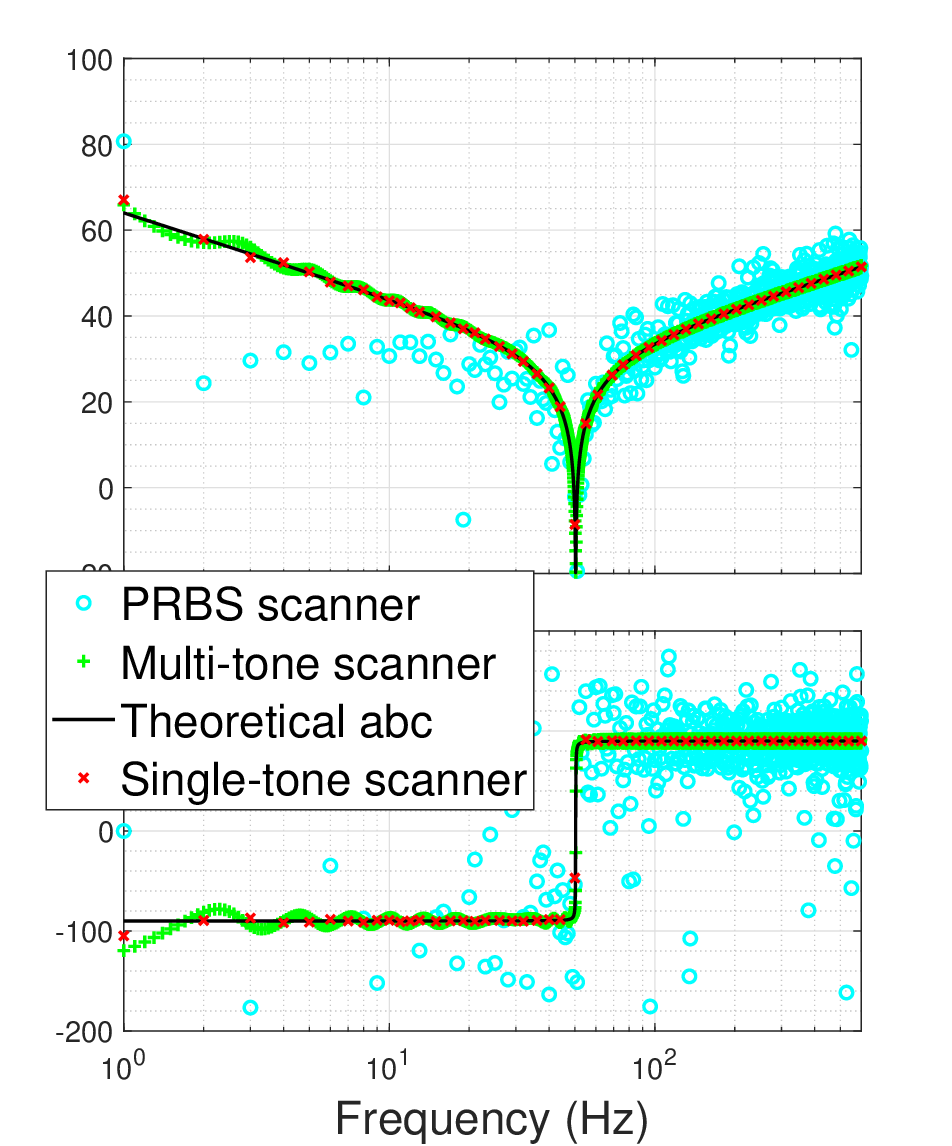}
        \vspace{-5mm}
        \caption{\( Z_{abc} \) with current perturbation.}
        \label{fig:RLCb}
    \end{subfigure}
    \vspace{-1mm}
    \caption{$abc$ frequency response of a series \( RLC \) load.}
    \label{fig:RLC}
    \vspace{-5mm}
\end{figure}

A balanced series \( RLC \) load (\( R = 0.1\,\Omega \), \( L = 0.1\,\mathrm{H} \), \( C = 100\,\mu\mathrm{F} \)) was analyzed using the scheme in Fig.~\ref{fig:RLCloadscheme} and various perturbation signals. The observation time was 3~s with a 1~s window, a frequency band of 1--600~Hz (50 logarithmic points), and a simulation step of \( \Delta t = 10\,\mu\mathrm{s} \). Validation compared the identified response with the theoretical admittance and impedance. Single-tone signals yielded the highest accuracy. For multi-tone and PRBS signals, the frequency band must be limited to ensure proper identification; among these, multi-tone performed better without additional filtering. For high-frequency identification (\(>1000\,\mathrm{Hz} \)), single-tone perturbations are recommended. Consequently, multi-tone and PRBS signals were avoided in subsequent tests to capture high-frequency behavior. Fig.~\ref{fig:RLC} shows the identified admittance and impedance matrices. For clarity, only one diagonal element is presented, as off-diagonal terms exhibit negligible numerical noise due to circuit symmetry.

\vspace{-4mm}

\subsection{PI Section Scanning}

\begin{figure}[t!]
    \centering
    \begin{subfigure}[b]{0.491\columnwidth}
        \centering
        \includegraphics[width=\textwidth]{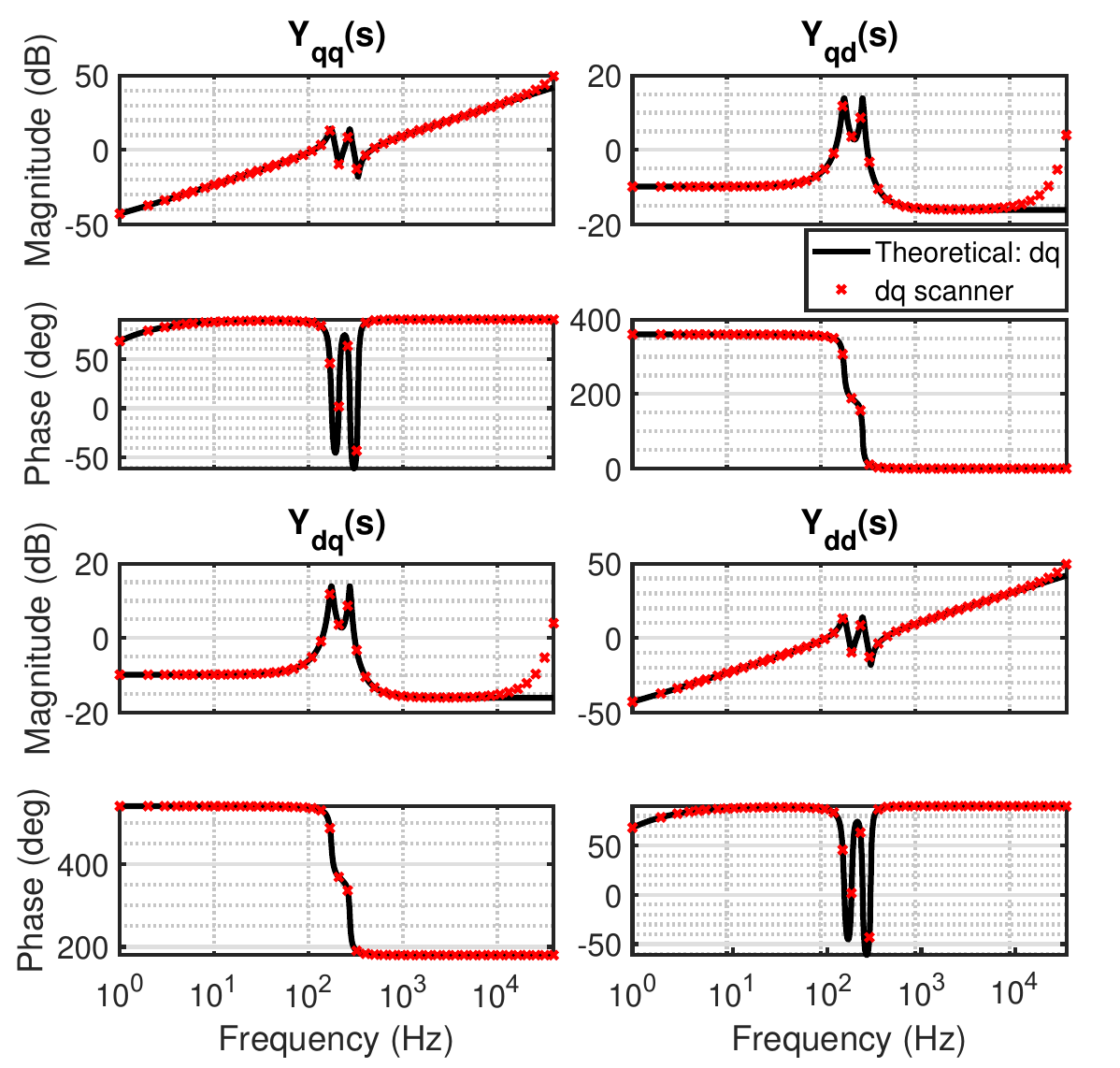}
        \vspace{-6mm}
        \caption{\( \mathbf{Y}_{dq} \) with voltage perturbation.}
        \label{fig:RLCa}
    \end{subfigure}
    \hfill
    \begin{subfigure}[b]{0.491\columnwidth}
        \centering
        \includegraphics[width=\textwidth]{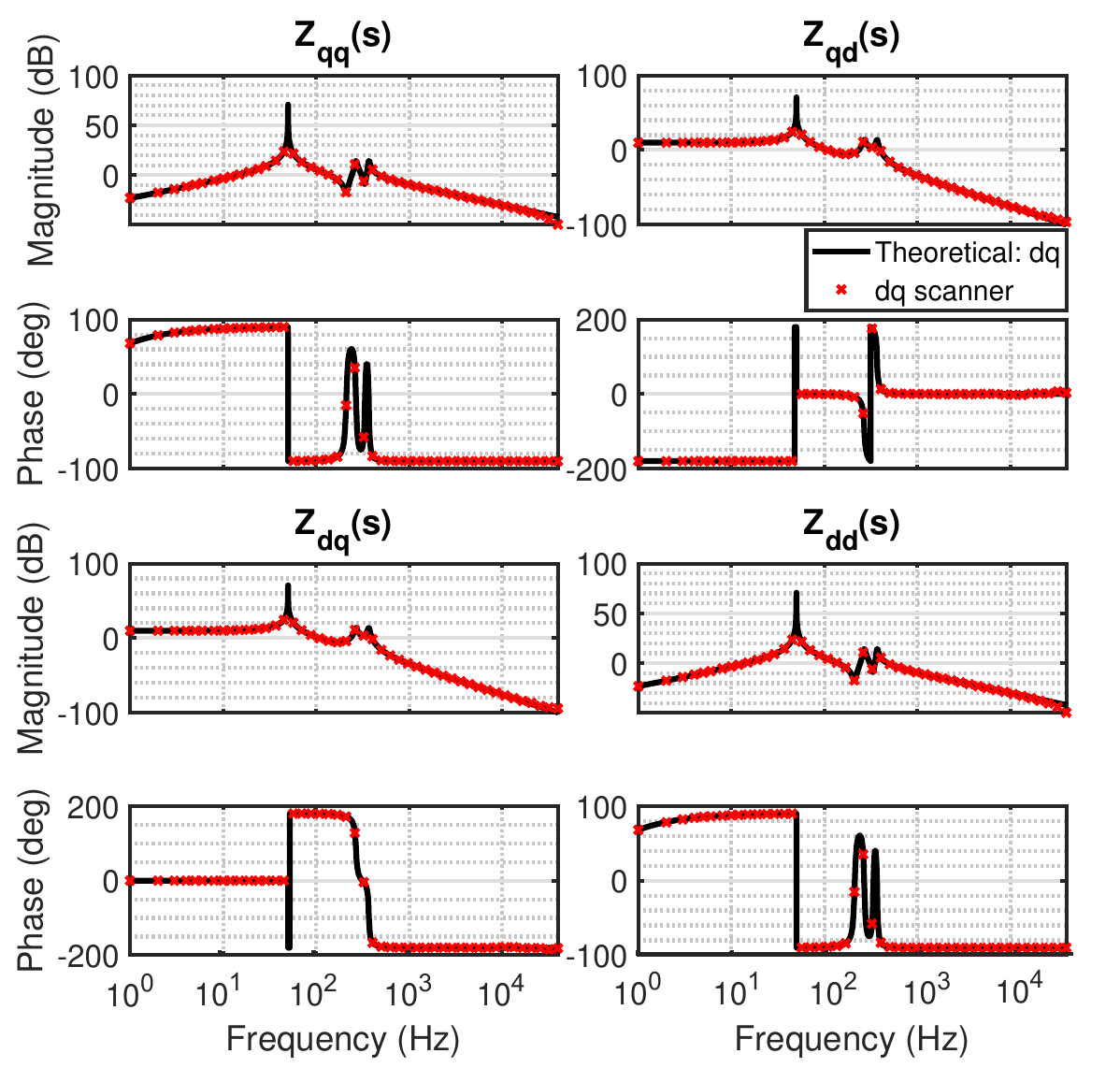}
        \vspace{-6mm}
        \caption{\( \mathbf{Z}_{dq} \) with current perturbation.}
        \label{fig:RLCb}
    \end{subfigure}
    \vspace{-6mm}
    \caption{Frequency response of a PI section in the \( dq \) frame.}
    \label{fig:PIdq0}
    \vspace{-3mm}
\end{figure}

To validate the scanner in the \( dq0 \) frame, a balanced PI section was analyzed using the configuration in Fig.~\ref{fig:PIsectionscheme}. Results were compared with the theoretical state-space response under open-circuit conditions, excluding the zero-sequence component. Parameters were \( R = 0.1\,\Omega \), \( L = 1\,\mathrm{mH} \), \( C = 1000\,\mu\mathrm{F} \), and a high-resistance load of \( R = 1\,\text{M}\Omega \). The simulation used a 5~s observation time, a 2~s window, and a time step of \( \Delta t = 10\,\mu\mathrm{s} \). Fig.~\ref{fig:PIdq0} shows the identified admittance and impedance matrices in the \( dq \) frame using single-tone perturbation. The results confirm high accuracy for both voltage and current perturbation strategies, validating the methodology in the \( dq0 \) domain.

\vspace{-4mm}

\subsection{Frequency-Dependent Line Scanning}

\begin{figure}[t!]
    \centering
    \begin{subfigure}[b]{0.491\columnwidth}
        \centering
        \includegraphics[width=\textwidth]{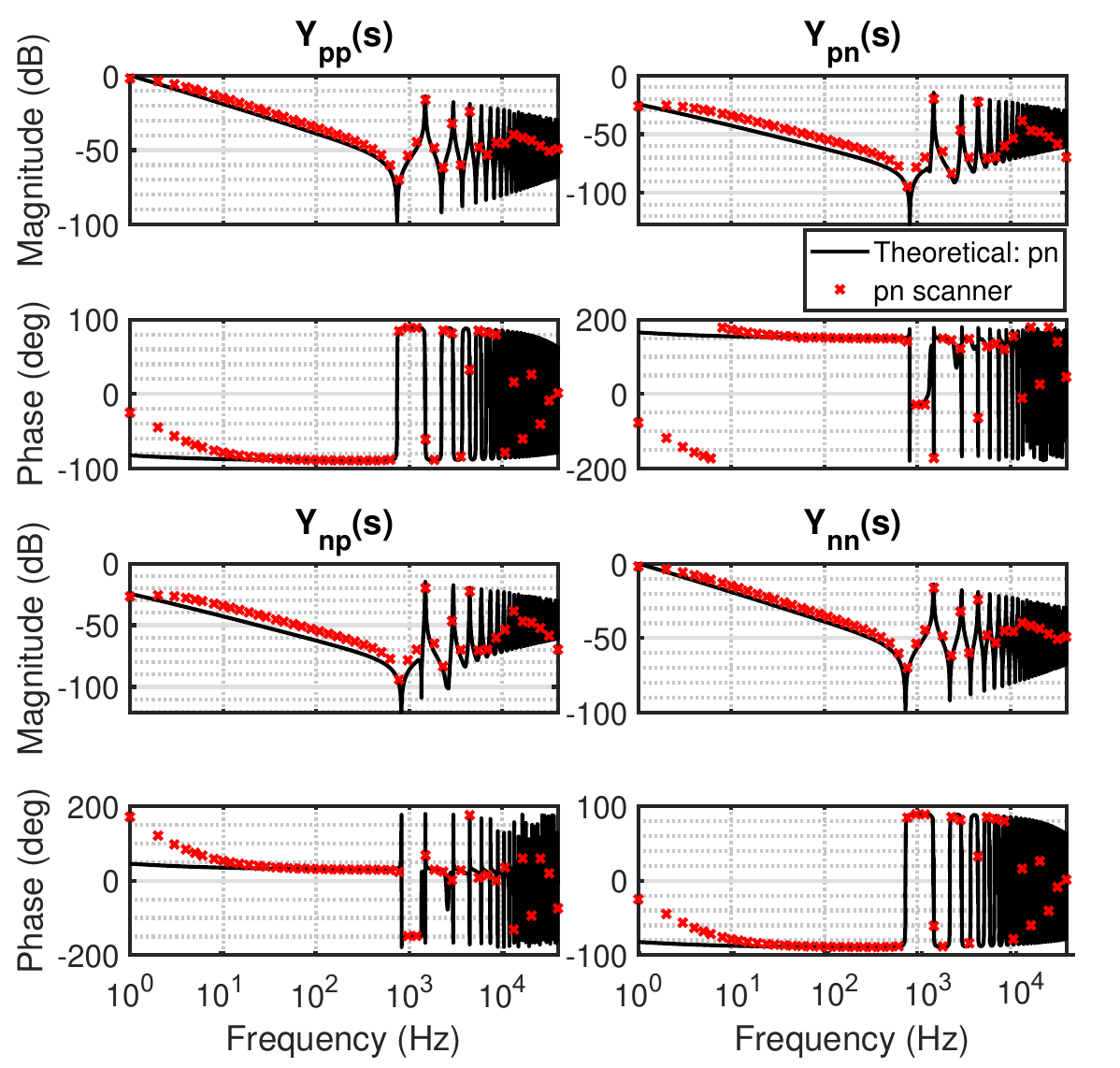}
        \vspace{-6mm}
        \caption{\( \mathbf{Y}_{pn} \) with voltage perturbation.}
        \label{fig:FDline_a}
    \end{subfigure}
    \hfill
    \begin{subfigure}[b]{0.491\columnwidth}
        \centering
        \includegraphics[width=\textwidth]{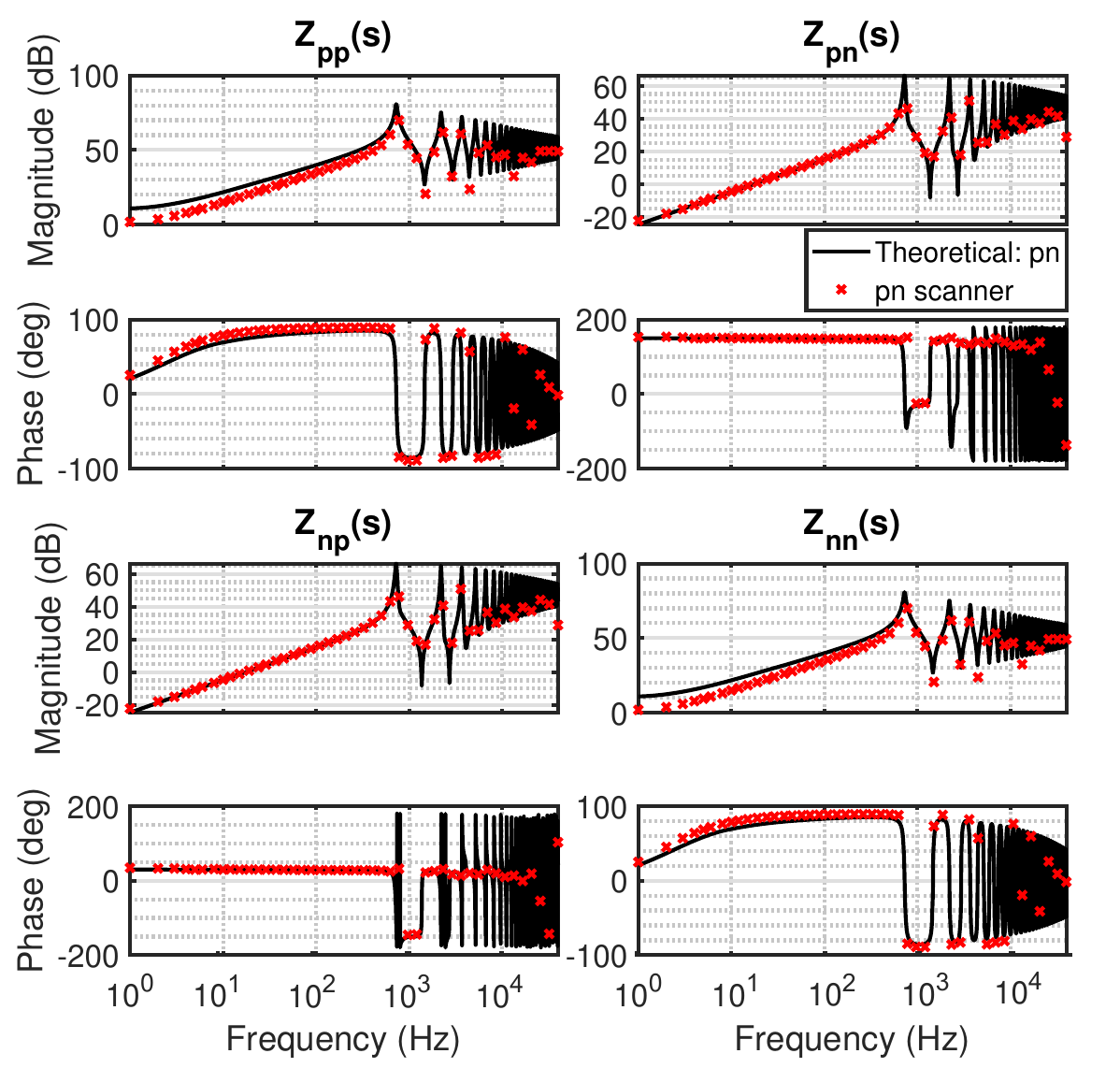}
        \vspace{-6mm}
        \caption{\( \mathbf{Z}_{pn} \) with current perturbation.}
        \label{fig:FDline_b}
    \end{subfigure}
    \vspace{-6mm}
    \caption{ $0pn$ Frequency response of a frequency-dependent line.}
    \label{fig:0pnFDline}
    \vspace{-5mm}
\end{figure}

To validate the scanner in the \( 0pn \) reference frame, a 100~km balanced overhead transmission line was analyzed using the parameters provided in \cite{paperJean}. The impedance and admittance matrices were validated against the theoretical two-port model response described in \cite{paperAbner}. The identified frequency responses are shown in Fig.~\ref{fig:0pnFDline}, where the \( pn \) sequence couplings are accurately captured. Results for the zero-sequence are omitted for simplicity, but exhibit comparable accuracy.

\vspace{-4mm}

\subsection{Grid-Following and Grid-Forming Converters}

\begin{figure}[t!]
    \centering
    \begin{subfigure}[b]{0.6\columnwidth}
        \centering
        \includegraphics[width=\textwidth]{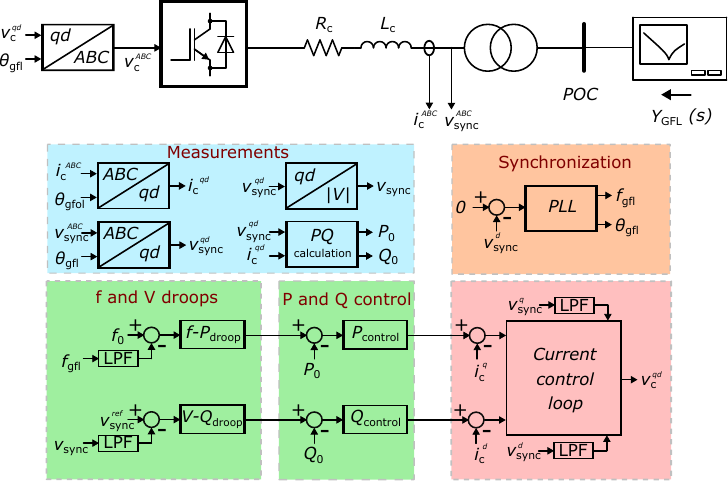}
        \vspace{-6mm}
        \caption{Grid-following.}
        \label{fig:GFLc}
    \end{subfigure}
    \vspace{0.5em}
    \begin{subfigure}[b]{0.6\columnwidth}
        \centering
        \includegraphics[width=\textwidth]{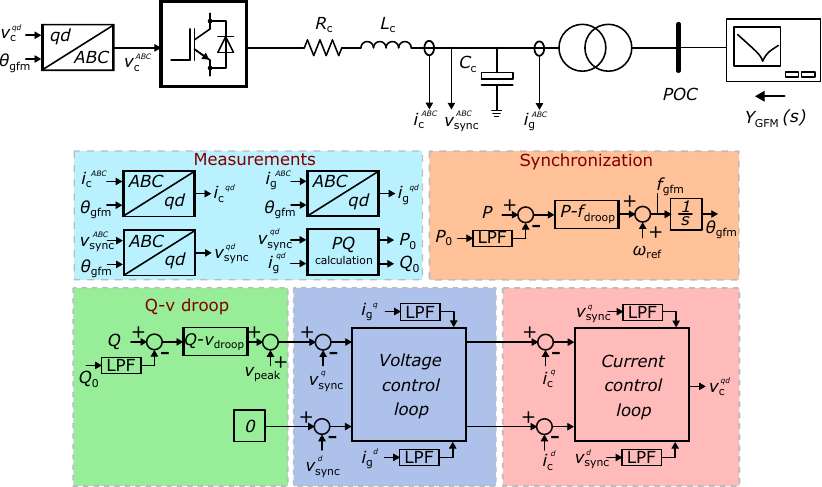}
        \vspace{-6mm}
        \caption{Grid-forming.}
        \label{fig:GFMc}
    \end{subfigure}
    \vspace{-4mm}
    \caption{Voltage source converter control schemes.}
    \label{fig:VSCscheme}
    \vspace{-3mm}
\end{figure}

To validate the FD scanning tool in modern power system applications, voltage source converter (VSC) models operating in grid-following (GFL) and grid-forming (GFM) modes were implemented, as shown in Fig.~\ref{fig:VSCscheme}. Parameters for both configurations are listed in Table~\ref{tab:table1}. Responses were validated against the linearized state-space representation derived using the open-source tool in \cite{STAMP}. A conventional PI-based PQ control scheme was employed, including voltage and current loops and a Phase-Locked Loop (PLL), following \cite{statespacebuilding}.

\begin{table}[t!]
\scriptsize
\setlength{\tabcolsep}{2pt}
\renewcommand{\arraystretch}{0.85}
\centering
\caption{Parameters for GFL and GFM converters.}
\label{tab:table1}
\vspace{-2mm}
\begin{tabular}{|c|c|c|c|c|}
\hline
\textbf{Param.} & \textbf{GFL} & \textbf{Unit} & \textbf{GFM} & \textbf{Unit} \\ \hline
$S_{base}$      & 2.750        & MVA           & 2.750        & MVA           \\ \hline
$V_{base}$      & 690.0        & V             & 690.0        & V             \\ \hline
$V_{DC}$        & 2000         & V             & 2000         & V             \\ \hline
$R_{C}$         & 0.034        & $\Omega$      & 0.034        & $\Omega$      \\ \hline
$L_{C}$         & 43.00        & mH            & 82.00        & mH            \\ \hline
$C_{C}$         & --           & --            & 9.193        & $\mu$F        \\ \hline
$R_{trans}$     & 0.003        & $\Omega$      & 0.003        & $\Omega$      \\ \hline
$L_{trans}$     & 4.300        & mH            & 8.300        & mH            \\ \hline
$k_{i_{PLL}}$   & 568.1        & pu            & --           & --            \\ \hline
$k_{p_{PLL}}$   & 1.420        & pu            & --           & --            \\ \hline
$k_{i_{CCL}}$   & 0.866        & pu            & 34.62        & pu            \\ \hline
$k_{p_{CCL}}$   & 1.378        & pu            & 0.083        & pu            \\ \hline
$k_{i_{PQ}}$    & 1.000        & pu            & --           & --            \\ \hline
$k_{p_{PQ}}$    & 0.100        & pu            & --           & --            \\ \hline
$\tau_{PQ}$     & 50.00        & ms            & --           & --            \\ \hline
$k_{i_{VCL}}$   & --           & --            & 420.0        & pu            \\ \hline
$k_{p_{VCL}}$   & --           & --            & 40.79        & pu            \\ \hline
$\omega_{PQ}$   & --           & --            & 25.00        & rad/s         \\ \hline
$\tau_{VI}$     & --           & --            & 10.00        & ms            \\ \hline
$k_{P}$         & 0.175        & pu            & 0.138        & pu/pu         \\ \hline
$k_{Q}$         & 0.073        & pu            & 0.275        & pu/pu         \\ \hline
\end{tabular}
\vspace{-4mm}
\end{table}

\begin{figure}[t!]
    \centering
    \begin{subfigure}[b]{0.491\columnwidth}
        \centering
        \includegraphics[width=\textwidth]{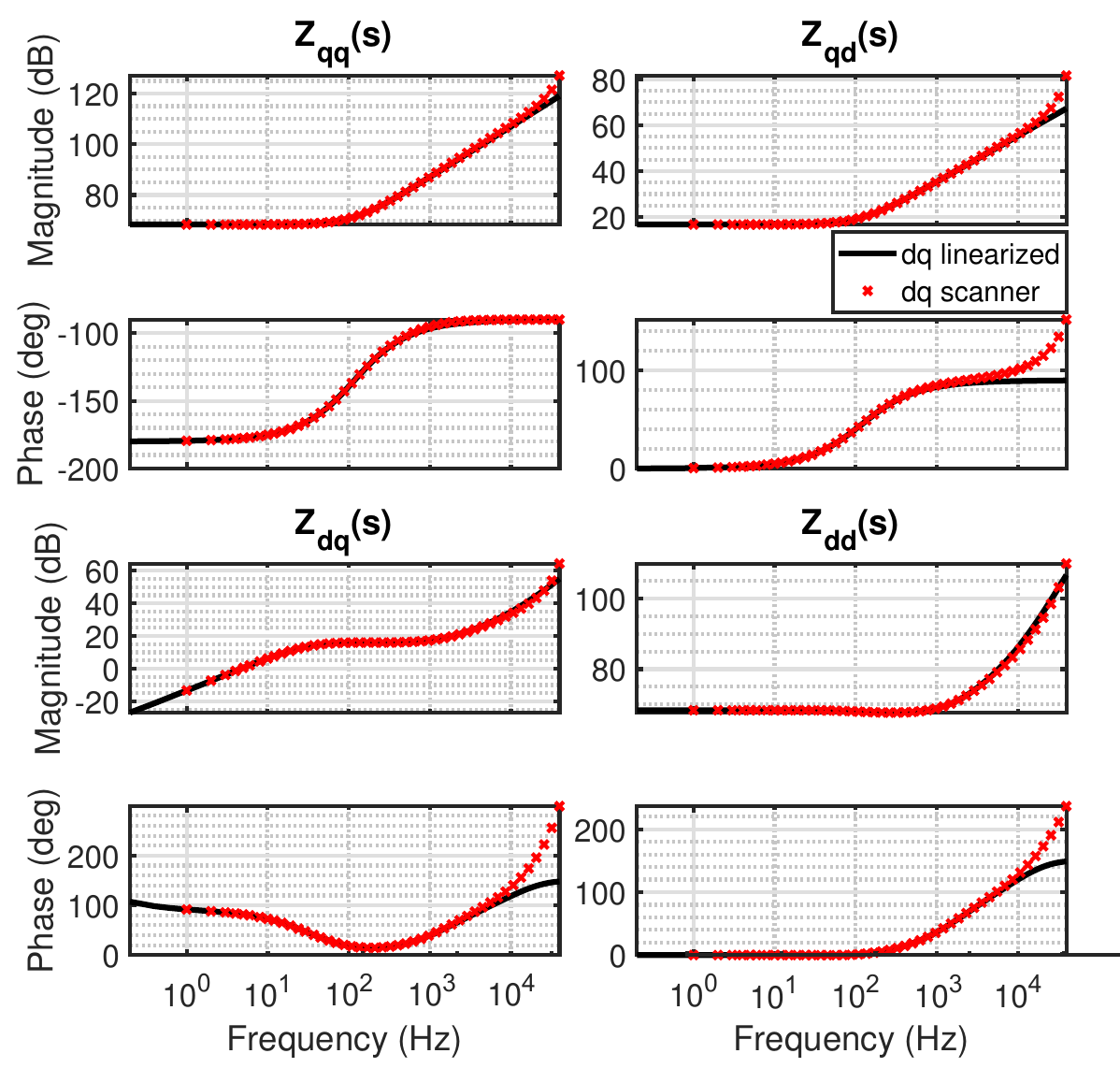}
        \vspace{-6mm}
        \caption{Grid-following.}
        \label{fig:GFLc}
    \end{subfigure}
    \vspace{0.5em}
    \begin{subfigure}[b]{0.491\columnwidth}
        \centering
        \includegraphics[width=\textwidth]{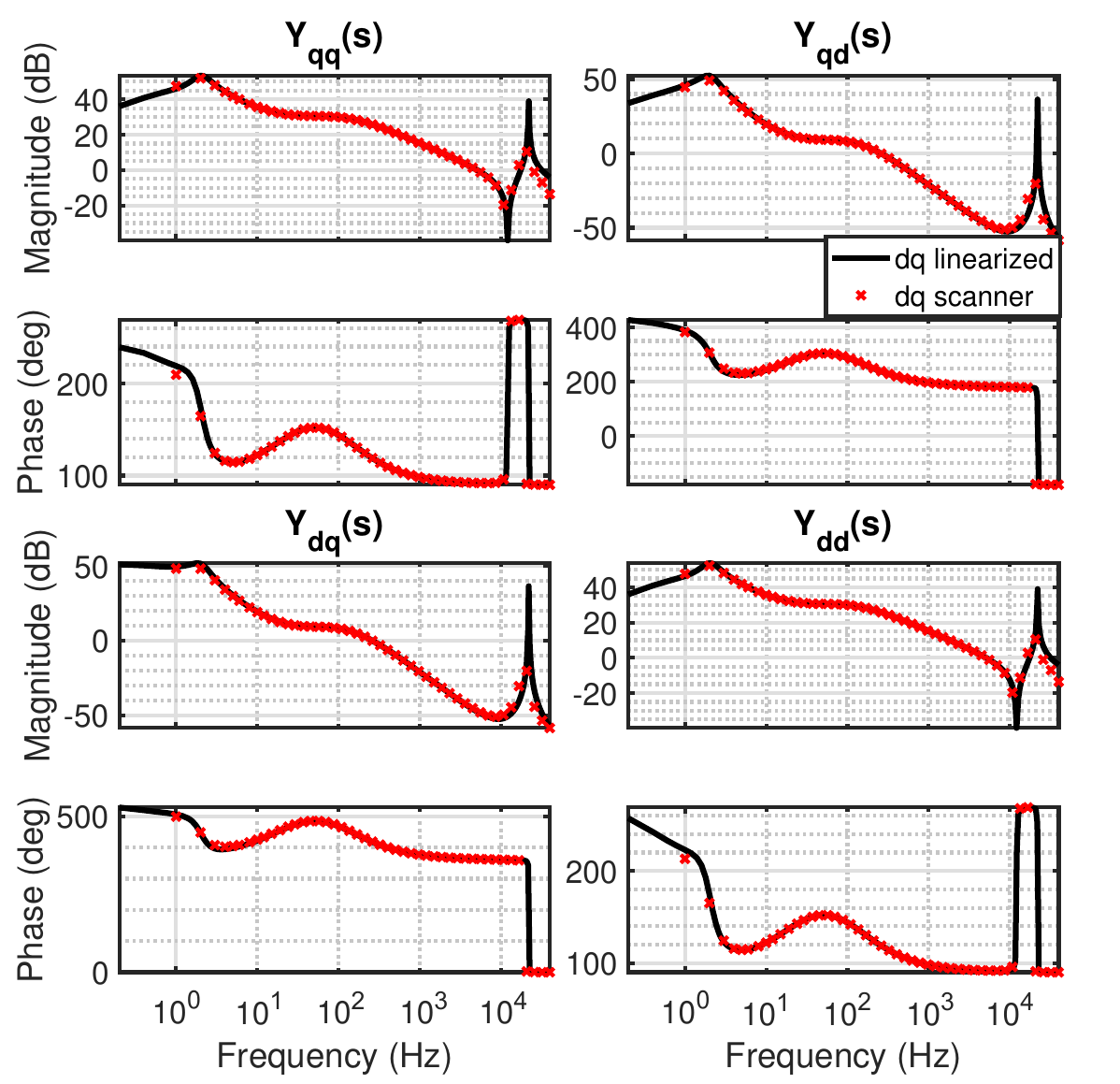}
        \vspace{-6mm}
        \caption{Grid-forming.}
        \label{fig:GFMc}
    \end{subfigure}
    \vspace{-8mm}
    \caption{Frequency response for converter control schemes.}
    \label{fig:dq0GF}
    \vspace{-5mm}
\end{figure}

The impedance and admittance frequency responses for VSC-GFL and VSC-GFM are presented in Fig.~\ref{fig:dq0GF}. Both exhibit high identification accuracy, with measured responses closely matching the linearized models at the operating point (\( P = 0.5 \), \( Q = 0.0 \,\text{pu} \)). The average scanning time was 38.6~s.

\vspace{-3mm}

\section{Stability and Interactions Assessment}

The SIaD-Tool integrates multiple modules for automatic stability and interaction assessment, comprising four distinct methodologies:  
B) the GNC, C) oscillation mode and participation factor analysis, D) stability margin evaluation via phase margins, and E) passivity analysis.  

The following subsection introduces the general framework of impedance-based analysis, which enables the application of all four methodologies implemented in the tool.

\vspace{-4mm}

\subsection{Impedance-Based Analysis in Modern Power Systems}

\begin{figure}[t!]
 \centering
\includegraphics[width=0.6\columnwidth]{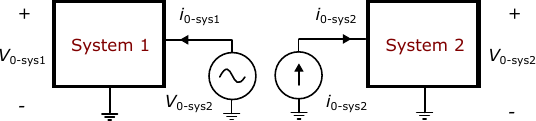}
 \vspace{-1mm}
 \caption{System equivalent decoupling scheme.}
 \label{fig:openloop}
 \vspace{-3mm}
\end{figure}

\begin{figure}[t!]
    \centering
\includegraphics[width=0.6\columnwidth]{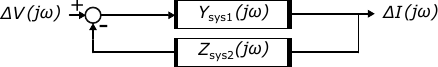}
    \vspace{-1mm}
    \caption{Closed-loop representation of the small-signal model.}
    \label{fig:openloop2}
    \vspace{-5mm}
\end{figure}

Once the admittance or impedance matrices of both subsystems—typically a power electronic converter and the grid or its equivalent—are identified in the FD, they can be modeled as shown in Figs.~\ref{fig:openloop} and \ref{fig:openloop2}, following the small-signal impedance ratio and closed-loop analysis in the FD \cite{papernyquist}.

The closed-loop transfer function is given by:
\begin{equation}\label{eq:closed_loop}
    \mathbf{T}(j\omega) = \left[ \mathbf{I} + \mathbf{Y}_{\text{sys1}}(j\omega) \, \mathbf{Z}_{\text{sys2}}(j\omega) \right]^{-1} \mathbf{Y}_{\text{sys1}}(j\omega) ,
\end{equation}
where the open-loop transfer function matrix is defined as:
\begin{equation}\label{eq:loop_transfer}
    \mathbf{L}(j\omega) = \mathbf{Y}_{\text{sys1}}(j\omega) \, \mathbf{Z}_{\text{sys2}}(j\omega).
\end{equation}

\vspace{-6mm}

\subsection{Generalized Nyquist Criterion (GNC)}

For Multiple-Input Multiple-Output (MIMO) systems, the GNC offers a graphical approach to assess closed-loop stability by analyzing the open-loop frequency response. SIaD-Tool implements this criterion in two ways: (i) by extracting eigenvalue trajectories of the loop transfer matrix, and (ii) by computing its determinant for screening. The determinant-based formulation is given by \cite{papernyquist}:
\begin{equation}\label{eq:GNC_loci}
    \det(\mathbf{I} + \mathbf{L}(j\omega)) = \prod_{i}(1 + \lambda_i(j\omega)),
\end{equation}
where \( \lambda_i(j\omega) \) denotes the \( i \)-th eigenvalue of the loop transfer matrix over \( j\omega \in [-\infty,\infty] \). Stability assessment follows \textbf{Definition 1} and the criteria in \cite{papernyquist,papernyquist2}.

\noindent\textbf{Definition 1 (GNC for Stability):}  
\textit{If the open-loop transfer function \(\mathbf{Y}_{\text{sys1}}(j\omega) \mathbf{Z}_{\text{sys2}}(j\omega) \) has no poles in the right-half complex plane (RHP), then the closed-loop system is stable if and only if none of the eigenvalue trajectories \( \lambda_i(j \omega) \) encircle the critical point \((-1, j0)\) in the Nyquist plot.}

In addition to stability verification, the GNC allows computing a stability margin metric following \cite{papernyquist}. The Nyquist Stability Margin (NSM) is the minimum distance between the Nyquist locus and the critical point $(-1,0)$ in the frequency range $\Omega$, indicating robustness: smaller values imply proximity to instability under gain or phase variations:
\begin{equation}
    \text{NSM} = \min_{\omega \in \Omega} \; \bigg( \min_{i \in \{1,\dots,x\}} \big| 1 + \lambda_i(j\omega) \big| \bigg).
\end{equation}
\vspace{-9mm}
\subsection{Impedance Mode Analysis}

To assess harmonic resonance interactions, the eigenvalues of the closed-loop admittance matrix are extracted for modal analysis, as proposed in \cite{impedance_modal}. This identifies critical modes at specific frequencies and determines which subsystem dominates each mode. Here, modal decomposition is applied:
\begin{equation}\label{eq:EVD_closed}
    \mathbf{Y}_{\text{sys}}(j\omega) = \boldsymbol{\Upsilon}(j\omega)\,\boldsymbol{\Lambda}(j\omega)\,\boldsymbol{\Phi}(j\omega),
\end{equation}
where \( \boldsymbol{\Upsilon}(j\omega) \) and \( \boldsymbol{\Phi}(j\omega) \) are right and left eigenvector matrices, and \( \boldsymbol{\Lambda}(j\omega) \) is the eigenvalue matrix.

The parallel admittance matrix is defined as \cite{impedance_modal}:
\begin{equation}\label{eq:Ysys}
    \mathbf{Y}_{\text{sys}}(j\omega) = \mathbf{Y}_{\text{sys1}}(j\omega) + \big[\mathbf{Z}_{\text{sys2}}(j\omega)\big]^{-1},
\end{equation}
and the modal impedance matrix is:
{\small
\begin{equation}\label{eq:impedance_modal}
    \mathbf{Z}_m(j\omega) = \big[\boldsymbol{\Lambda}(j\omega)\big]^{-1} =
    \begin{bmatrix}
        \lambda_1^{-1} & 0 & \cdots & 0 \\
        0 & \lambda_2^{-1} & \cdots & 0 \\
        \vdots & \vdots & \ddots & \vdots \\
        0 & 0 & \cdots & \lambda_x^{-1}
    \end{bmatrix},
\end{equation}}
where \( x \) is the dimension of the reference frame (\( abc \), \( dq0 \), or \( 0pn \)). When \( |\lambda_x(j\omega)| \to 0 \), then \( |Z_m(j\omega)| \to \infty \), indicating high modal impedance \cite{impedance_modal}. The dominant mode is the eigenvalue closest to the origin (smallest magnitude). Participation factors quantify each input influence on mode \(i\):
\begin{equation}
    P_{ki}(j\omega) = \frac{|\Upsilon_{ki}(j\omega)\,\Phi_{ik}(j\omega)|}{\sum_{k=1}^{N}|\Upsilon_{ki}(j\omega)\,\Phi_{ik}(j\omega)|},
\end{equation}
where \( P_{ki}(j\omega) \) is the normalized participation factor (PF).

\vspace{-4mm}

\subsection{Phase Margin Stability}

Oscillation frequencies can be identified through the intersection points of the impedance magnitudes of both subsystems. The interaction between the sys1 and sys2 is evaluated according to \textbf{Definition 2} \cite{papernyquist,passivityandstability}.

\noindent\textbf{Definition 2 (Stability Condition Based on Phase Margin):}  
\textit{The stability of the interconnected system can be assessed by analyzing the interaction between \( \mathbf{Y}_{\text{sys1}}(j \omega) \) and \( \mathbf{Z}_{\text{sys2}}(j \omega) \) in the frame ($0pn$) or ($dq$). Specifically, the phase margin (PM) at the intersection frequency \( \omega_i \) is defined as:}
\begin{equation}\label{eq:pm}
\text{PM} = 180^\circ + \left[ \angle \mathbf{Y}_{\text{sys1}}^{-1}(j \omega_i) - \angle \mathbf{Z}_{\text{sys2}}(j \omega_i) \right].
\end{equation}
\textit{A large positive phase margin (\( \text{PM} > 0 \)) indicates that the closed-loop system is stable under the given operating conditions. Conversely, if the phase margin is negative or close to zero, oscillatory behavior may arise, potentially leading to instability in the overall system.}

\vspace{-4mm}

\subsection{Passivity-Based Stability Assessment}

Passivity theory asserts that a passive system is inherently stable. Thus, ensuring passivity in specific network sections—such as HVDC links, STATCOMs, or selected frequency ranges—can provide a sufficient (though not necessary) condition for stability \cite{paperJosep,passivityandstability}.

For an admittance representation, passivity holds if \cite{paperJosep}:
\begin{enumerate}
    \item \( \mathbf{Y}(j\omega) \) has no poles in the right-half plane (RHP), ensuring stability across all frequencies.
    \item \( \mathbf{Y}(j\omega) + \mathbf{Y}^{H}(j\omega) \geq 0 \), meaning the minimum eigenvalue \( \min(\boldsymbol{\lambda}(j\omega)) \) is positive for all \( \omega \).
\end{enumerate}

SIaD-Tool evaluates passivity for both identified subsystems, enabling determination of operating ranges where passivity—and thus stability—is guaranteed. This provides an additional robustness metric and supports control design in modern power systems.

\vspace{-3mm}

\section{Application and Study Cases}

In the following section, the SIaD-Tool is applied to several representative scenarios to perform a comprehensive stability and interaction assessment in modern power systems.
\vspace{-3mm}

\subsection{Case Study I: Grid-Connected Grid-Following}

Using the GFL control scheme in Fig.~\ref{fig:GFLc}, a Thevenin equivalent is connected to the PoS at the same base voltage, with base SCR = 3 and \( X/R = 3 \). Remaining parameters follow Table~\ref{tab:table1}. Simulation time is 8~s with a step size of \( \Delta t = 50\,\mu\text{s} \). The SCR is varied from 0.8 to 3—without saturation strategy—to determine stability limits via the GNC, representing a screening study for modern grids. Voltage perturbation with multi-tone signals in the \( dq \) frame is used, covering 1–600~Hz for rapid boundary identification.

\begin{figure}[t!]
    \centering
    \begin{subfigure}[b]{0.7\columnwidth}
        \centering
        \includegraphics[width=0.85\textwidth]{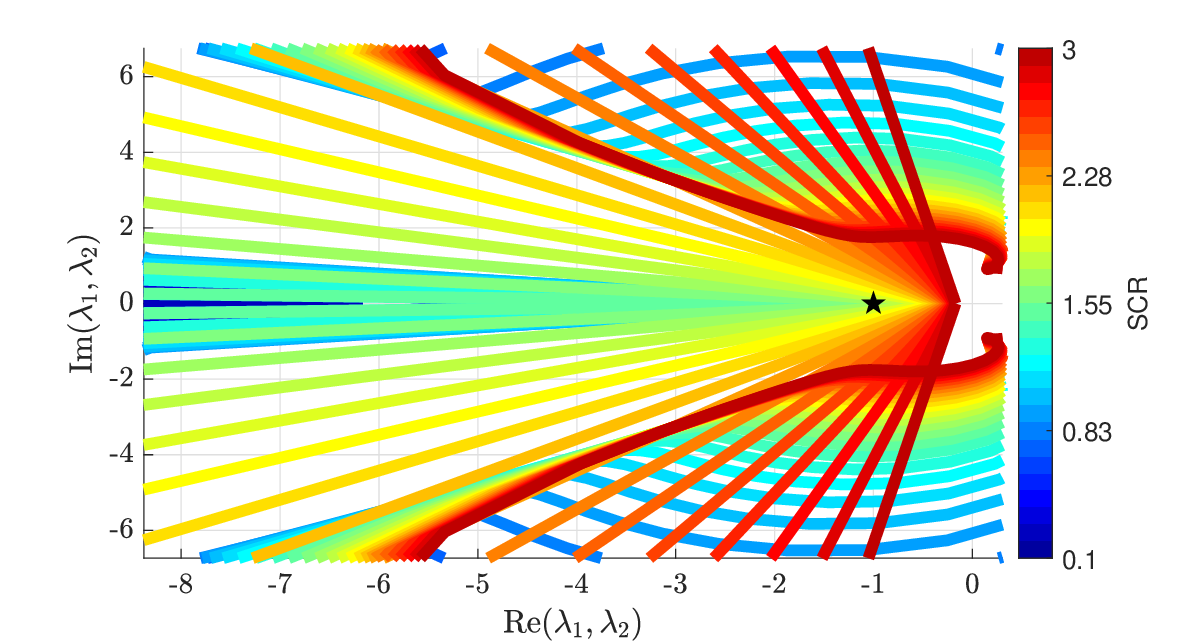}
        \vspace{-2mm}
        \caption{Screening via GNC.}
        \label{fig:GNC_SCR}
    \end{subfigure}
    \vspace{-4mm}
    \begin{subfigure}[b]{0.7\columnwidth}
        \centering
        \includegraphics[width=0.85\textwidth]{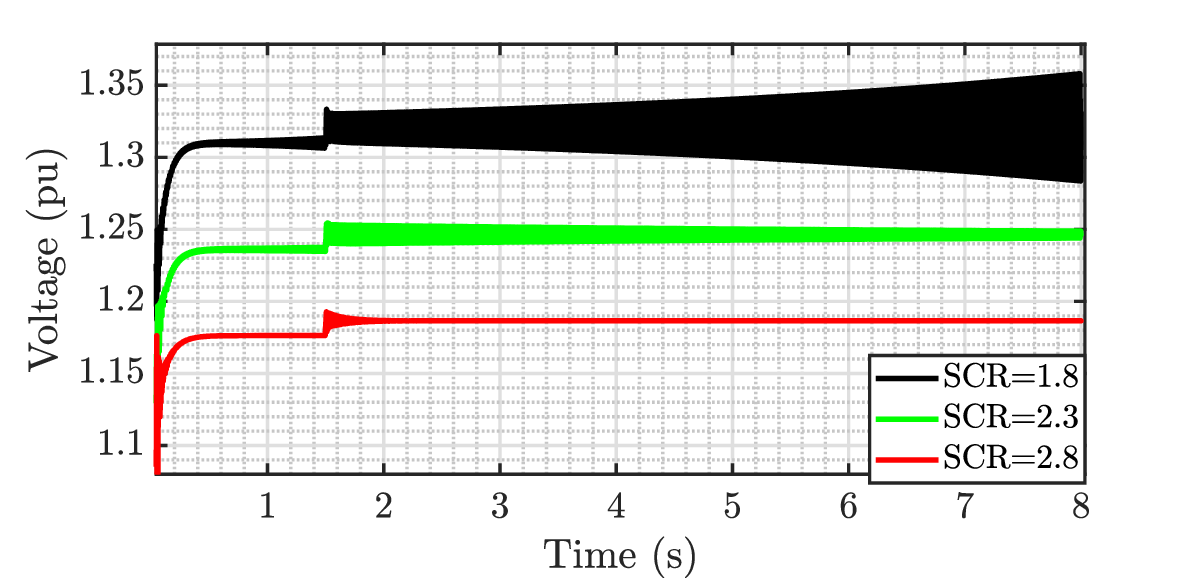}
        \vspace{-2mm}
        \caption{Time-domain simulations at POC.}
        \label{fig:TD_SCR}
    \end{subfigure}
    \vspace{3mm}
    \caption{Stability margin screening for varying SCR values.}
    \label{fig:SCR_var}
    \vspace{-3mm}
\end{figure}

Fig.~\ref{fig:SCR_var} shows the stability assessment. As seen in Fig.~\ref{fig:GNC_SCR}, instability occurs when SCR falls below 2.2. Eigenvalue trajectories in the \( dq \) frame indicate \( \lambda_1 \) remains outside \((-1,j0)\), while \( \lambda_2 \) encircles it as SCR drops below 2.1, going from a NSM of 1.49 to 0.05. Proximity to the critical point reduces stability margin, increasing susceptibility to instability. This is validated by a 1\% voltage step perturbation at three SCR values identified in the screening. Fig.~\ref{fig:TD_SCR} confirms that, despite increased voltage amplitude at lower SCR, the system remains stable for SCR $\geq$ 2.3, supporting the screening results.

\vspace{-4mm}

\subsection{Case Study II: HVAC Offshore Wind Power Plant}

\begin{figure}[t!]
\centering
\includegraphics[width=\columnwidth]{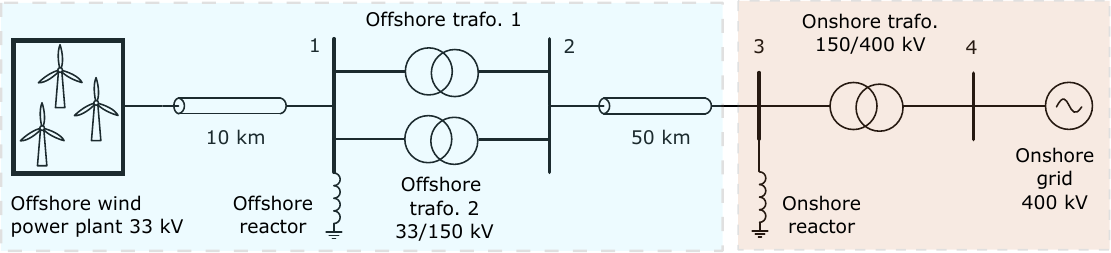}
\caption{HVAC offshore wind power plant system.}
\label{fig:owpp}
\vspace{-3.5mm}
\end{figure}

The SIaD-Tool methodology is applied to a 50~Hz HVAC offshore wind power plant (OWPP), shown in Fig.~\ref{fig:owpp}, to validate and analyze stability and interactions in a representative converter-based integration scenario. The complete Simulink model and data are available in \cite{SIaD_tool}.

Simulation parameters include \( T_{\text{init}} = 8 \)~s, frequency resolution of 1~Hz, time step \( \Delta t = 50\,\mu\text{s} \), and 113 logarithmic frequency points within \( [1, f_{\text{sampling}}/34] \), as per \eqref{eq:frule}. Single-tone voltage perturbation is used in the \( dq0 \) frame.

\begin{figure}[t!]
    \centering
    \vspace{-2mm}
    \begin{subfigure}[b]{0.491\columnwidth}
        \centering
        \includegraphics[width=\textwidth]{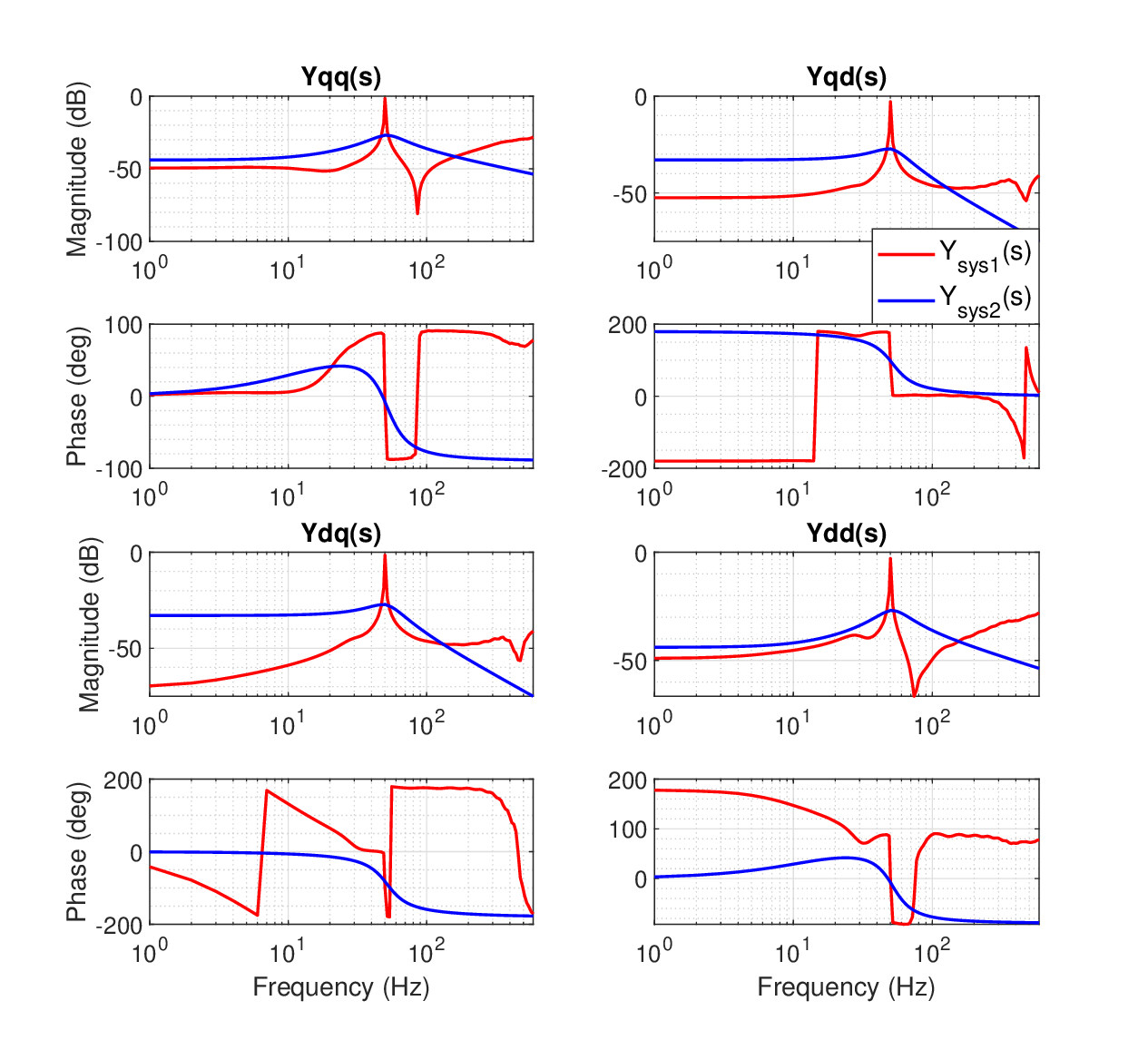}
        \vspace{-7mm}
        \caption{Stable case.}
    \end{subfigure}
    \vspace{-3mm}
    \begin{subfigure}[b]{0.491\columnwidth}
        \centering
        \includegraphics[width=\textwidth]{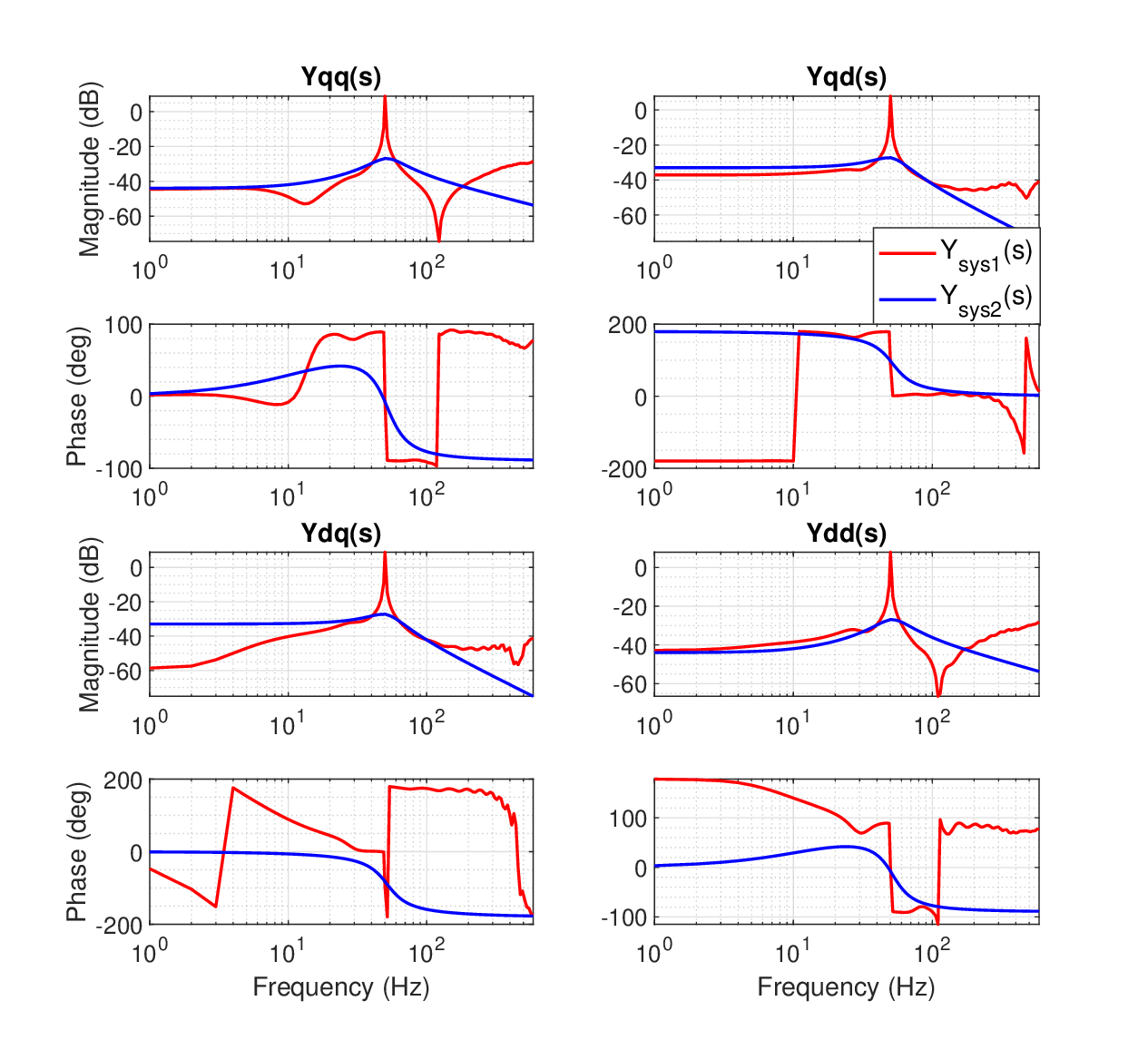}
        \vspace{-7mm}
        \caption{Unstable case.}
    \end{subfigure}
    \vspace{-3mm}
    \caption{Frequency response of the OWPP.}
    \label{fig:Zstable1}
    \vspace{-4mm}
\end{figure}

The study varies the onshore compensation reactor inductance from 0.9~H to 0.245~H, with OWPP operating at \( P = 0.9 \), \( Q = 0.1 \)~pu. Fig.~\ref{fig:Zstable1} compares frequency responses under stable and unstable conditions. In the stable case, PMs are mostly positive: \( 95.8^\circ \), \( 192^\circ \), \( 160^\circ \), and \( 95.2^\circ \) for \( Y_{qq}(s) \), \( Y_{qd}(s) \), \( Y_{dq}(s) \), and \( Y_{dd}(s) \), respectively, indicating an oscillatory point at 132~Hz in \( dq \)-coupling. Conversely, the unstable case shows a very negative PM in \( Y_{dq}(s) \) at 99~Hz.

\begin{figure}[t!]
    \centering
    \begin{subfigure}[b]{0.491\columnwidth}
        \centering
        \includegraphics[width=\textwidth]{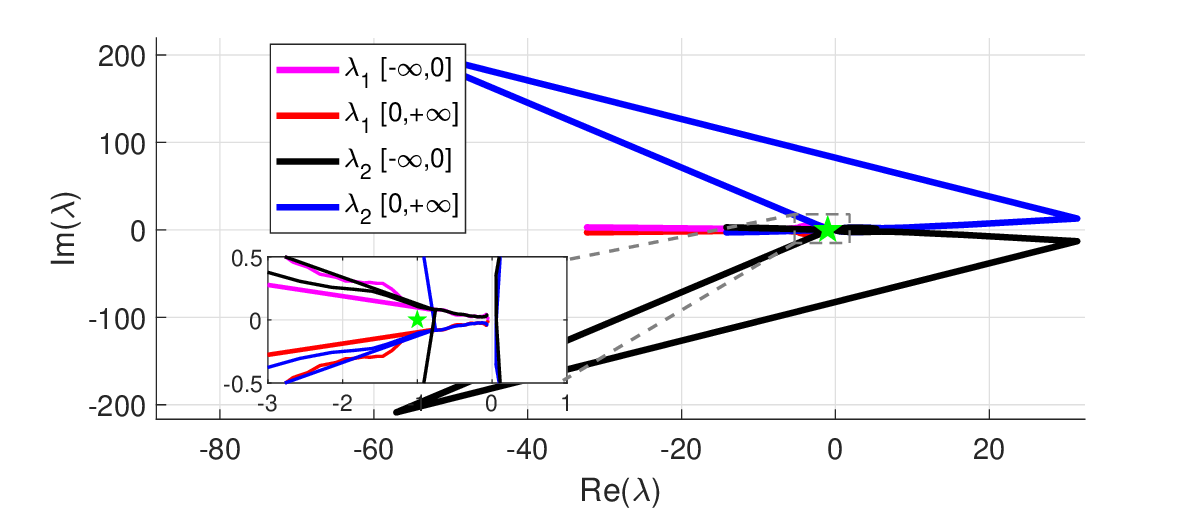}
        \vspace{-5mm}
        \caption{Stable case.}
    \end{subfigure}
    \vspace{0.5em}
    \begin{subfigure}[b]{0.491\columnwidth}
        \centering
        \includegraphics[width=\textwidth]{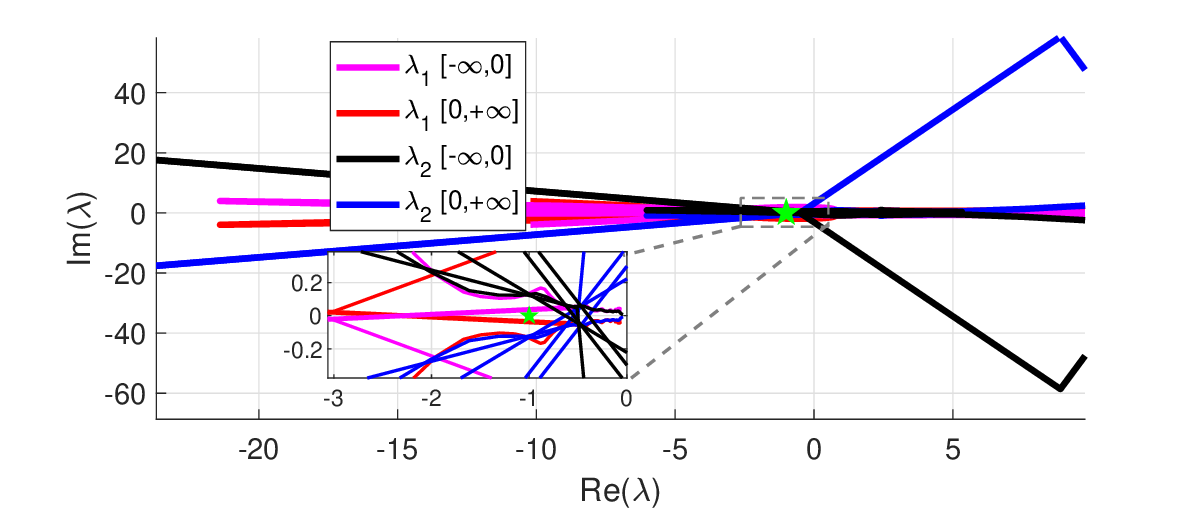}
        \vspace{-5mm}
        \caption{Unstable case.}
    \end{subfigure}
    \vspace{-7mm}
    \caption{Nyquist plots of the closed-loop response.}
    \label{fig:gnc1_sta}
    \vspace{-4mm}
\end{figure}
\begin{figure}[t!]
    \centering
    \begin{subfigure}[b]{0.491\columnwidth}
        \centering
        \includegraphics[width=\textwidth]{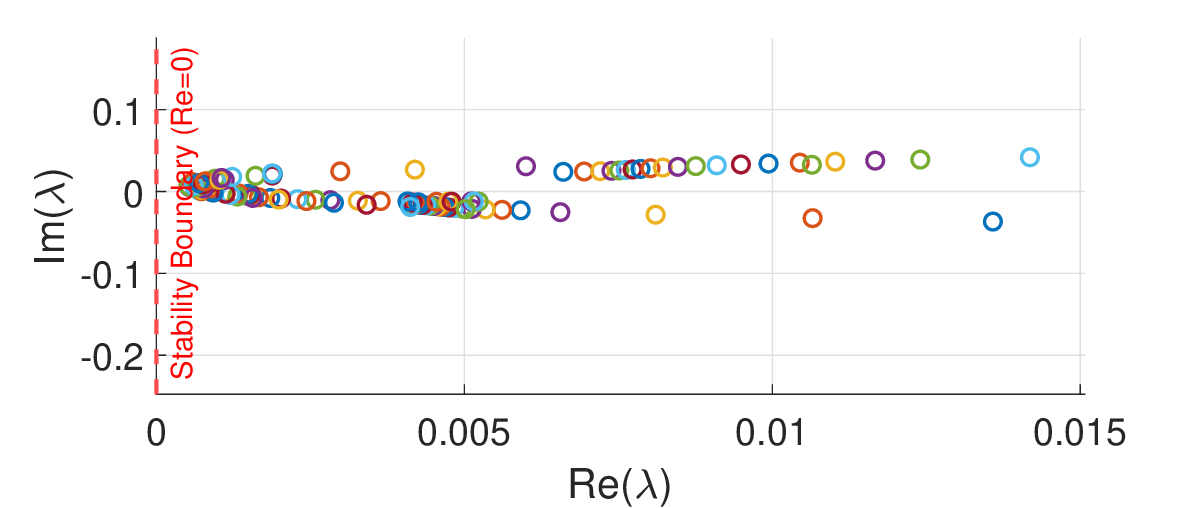}
        \vspace{-5mm}
        \caption{Stable case.}
    \end{subfigure}
    \vspace{0.5em}
    \begin{subfigure}[b]{0.491\columnwidth}
        \centering
        \includegraphics[width=\textwidth]{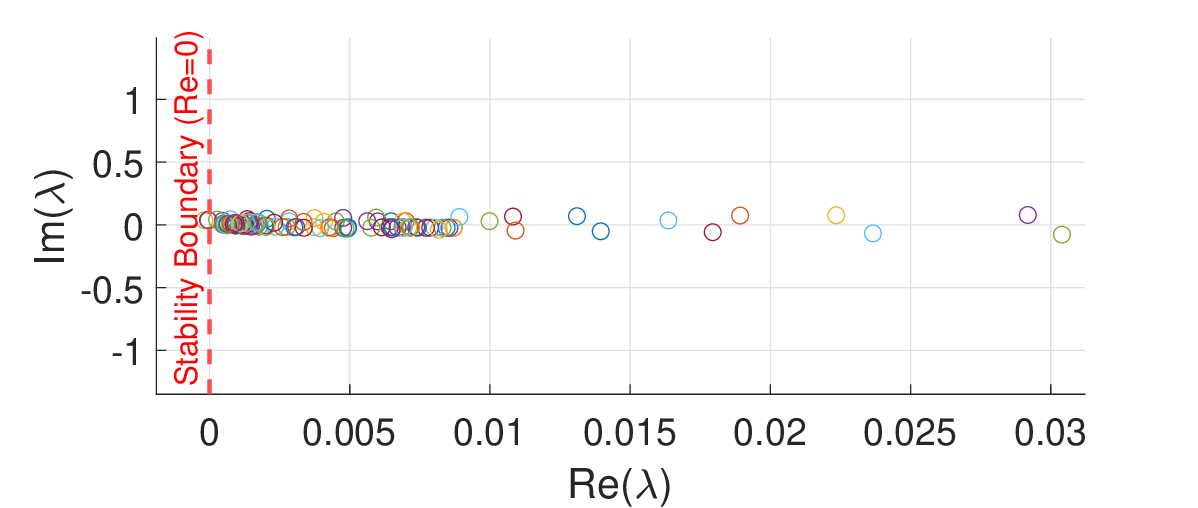}
        \vspace{-5mm}
        \caption{Unstable case.}
    \end{subfigure}
    \vspace{-8mm}
    \caption{Eigenvalue loci of $\mathbf{Y}_{\text{sys}}(j\omega)$ across the frequencies.}
    \label{fig:Zm_owpp}
    \vspace{-4mm}
\end{figure}
\begin{figure}[t!]
    \centering
    \begin{subfigure}[b]{0.491\columnwidth}
        \centering
        \includegraphics[width=\textwidth]{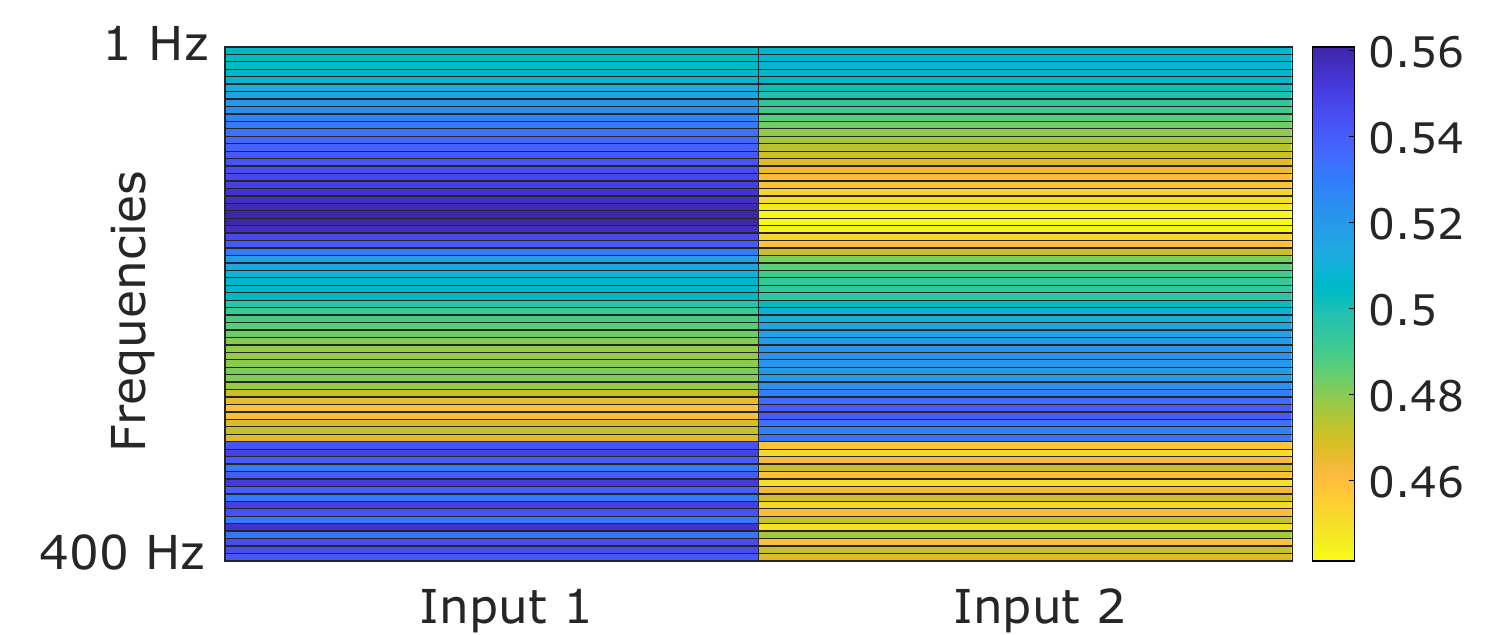}
        \vspace{-5mm}
        \caption{Stable case.}
    \end{subfigure}
    \vspace{0.5em}
    \begin{subfigure}[b]{0.491\columnwidth}
        \centering
        \includegraphics[width=\textwidth]{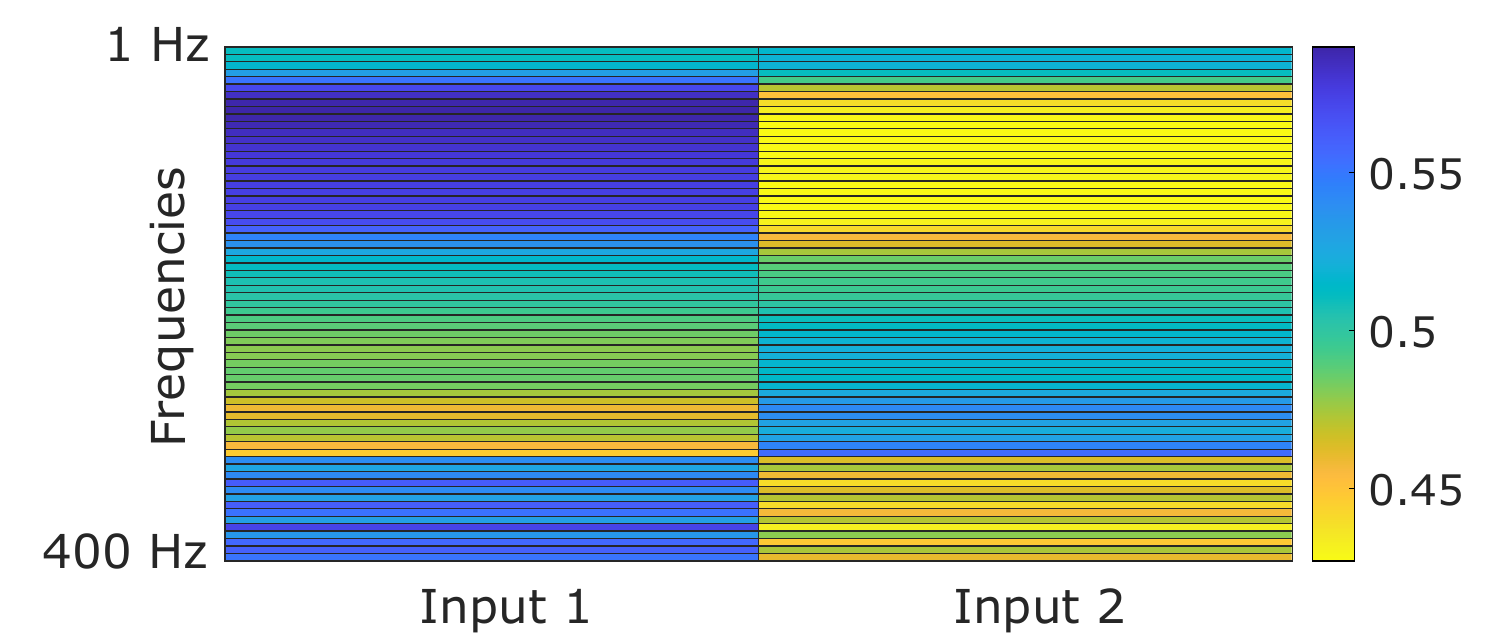}
        \vspace{-5mm}
        \caption{Unstable case.}
    \end{subfigure}
    \vspace{-8mm}
    \caption{Summary of participation factors of critical modes.}
    \label{fig:pf_owpp}
    \vspace{-5mm}
\end{figure}
\begin{figure}[t!]
\centering
\includegraphics[width=0.6\columnwidth]{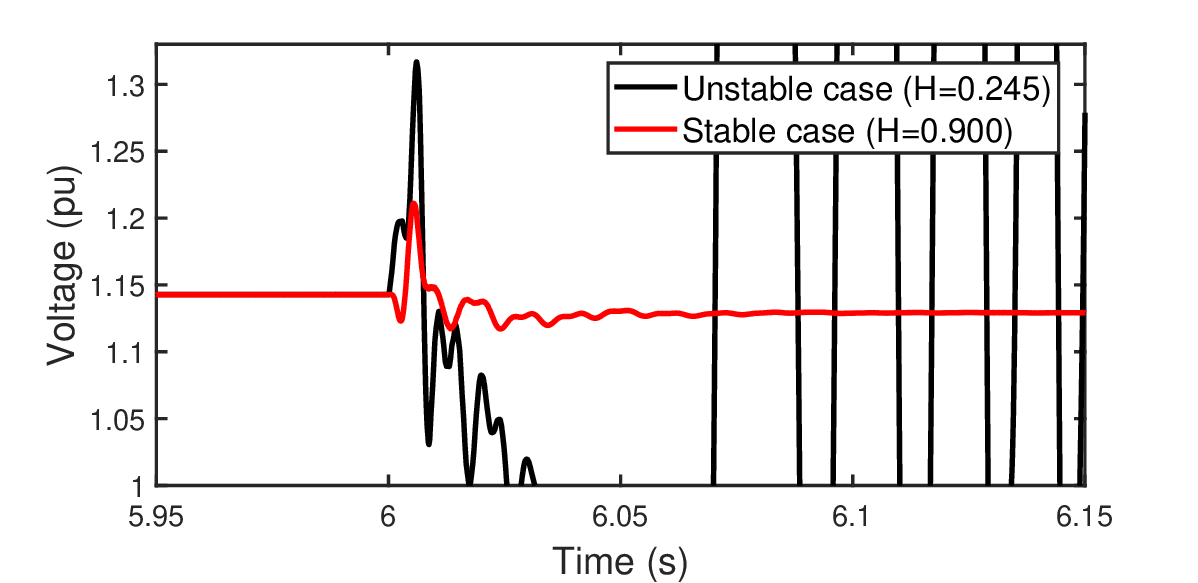}
\vspace{-2mm}
\caption{EMT transient validation for the OWPP.}
\label{fig:td_owpp}
\vspace{-6mm}
\end{figure}

Further stability evaluation using Nyquist plots in Fig.~\ref{fig:gnc1_sta} shows that, in the stable case, eigenvalue trajectories do not encircle the critical point with a NSM of 0.21. Conversely, in the unstable case, \( \lambda_1 \) encircles this point with a NSM of 0.08, confirming instability. In practice, graphical stability assessment via GNC is challenging; therefore, this article employs eigenvalue analysis to determine system stability. Figs.~\ref{fig:Zm_owpp} and \ref{fig:pf_owpp} present eigenvalues of \(\mathbf{Y}_{\text{sys}}(j\omega)\) and PFs for selected frequencies (1–400~Hz). In Fig.~\ref{fig:Zm_owpp}(a), stability holds as no eigenvalue crosses the negative side; however, reduced reactance causes crossings (Fig.~\ref{fig:Zm_owpp}(b)), corroborating GNC results. Fig.~\ref{fig:pf_owpp} illustrates frequency-dependent participation: at low frequencies, input~1 (OWPP-dominated) prevails near 9~Hz in the unstable case, while in the stable case dominance shifts toward 28~Hz. Beyond 40~Hz, participation remains nearly constant, quantitatively identifying sensitivity frequencies. Time-domain simulations in Fig.~\ref{fig:td_owpp} validate these findings, revealing an initial oscillation near 99~Hz after a 1\% voltage step—consistent with PM analysis. Modal analysis for the stable case identifies dominant modes at 103~Hz (grid-dominated) and 202~Hz (OWPP-dominated), indicating resonance susceptibility. In the unstable case, dominant modes occur at 118~Hz and 217~Hz, corresponding to the highest \(|Z_m|\) values. Finally, passivity analysis of open-loop responses highlights non-passive regions for the OWPP at 10–26~Hz, 50~Hz, 58–67~Hz, and 92–202~Hz. Under unstable conditions, these regions expand, increasing instability risk due to OWPP active-source behavior.
\vspace{-4mm}

\subsection{Case Study III: IEEE 9-Bus System}

\begin{figure}[t!]
\centering
\includegraphics[width=0.7\columnwidth]{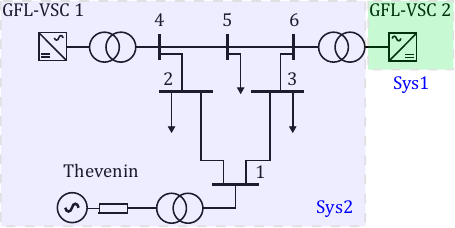}
\caption{Test network based on the IEEE 9-bus system.}
\label{fig:9bus}
\vspace{-2mm}
\end{figure}

The SIaD-Tool is applied to the IEEE 9-bus benchmark system to validate scalability and robustness in stability and interaction assessment for modern power systems. Fig.~\ref{fig:9bus} illustrates the network configuration, where two GFL converters are connected at buses 4 and 6 via transformers. The complete Simulink model, including all parameters and grid topology, is available in the repository \cite{SIaD_tool}.

The PQ operating points are set to \( [1.590, 0.065] \)~pu at bus 4 and \( [0.829, 0.000] \)~pu at bus 6. The SCR from the source is fixed, transmission lines are modeled as PI sections, and RL loads are connected in shunt. Evaluation uses an observation time of 8~s, a time step of \( \Delta t = 25\,\mu\text{s} \), and 100 logarithmic frequency points from 1~Hz to 1250~Hz, with voltage perturbation in the \( dq \)-frame.

SIaD-Tool is placed between GFL2 and the transformer before node 6 to analyze the interaction between the converter (Sys1) and the grid (Sys2). Using impedance modal analysis and phase margin methodology from \eqref{eq:impedance_modal} and \eqref{eq:pm}, modal impedances and participation factors are computed for each frequency. Dominant eigenvalues are identified, and system dominance is determined by the highest participation factor. Table~\ref{tab:table_sta1} summarizes the parallel resonance analysis.

\begin{table}[t!]
\centering
\caption{Summary of impedance analysis.}
\label{tab:table_sta1}
\begin{tabular}{|c|c|c|c|c|c|}
\hline
\textbf{f (Hz)} & \textbf{Dominant} & \( \lambda \) & \( |Z_m| \) & PF1 & PF2 \\ \hline
1.000 & Sys2 & 1.520 - j3.306 & 0.275 & 0.481 & 0.518 \\ \hline
5.000 & Sys1 & 1.592 - j3.493 & 0.260 & 0.530 & 0.471 \\ \hline
8.000 & Sys2 & -0.794 - j0.214 & 1.214 & 0.453 & 0.547 \\ \hline
30.00 & Sys2 & -0.573 + j0.711 & 1.094 & 0.411 & 0.687 \\ \hline
100.0 & Sys1 & 1.815 - j1.235 & 0.455 & 0.500 & 0.499 \\ \hline
1250 & Sys1 & 0.478 - j1.008 & 0.896 & 0.5001 & 0.4999 \\ \hline
\end{tabular}
\vspace{-4mm}
\end{table}

\begin{figure}[t!]
\centering
\includegraphics[width=0.75\columnwidth]{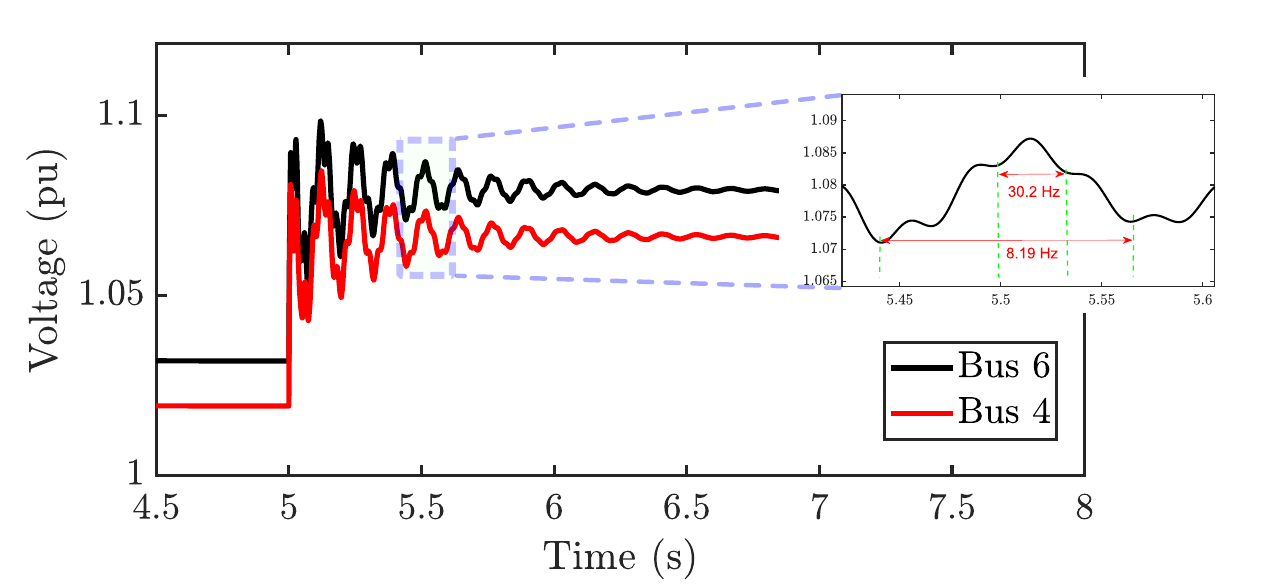}
\vspace{-2mm}
\caption{Transient validation for the IEEE 9-bus system.}
\label{fig:9bustimeresponse}
\vspace{-3mm}
\end{figure}

Two modal impedances with the highest magnitudes occur at 8~Hz and 30~Hz, with values of 1.214 and 1.094, respectively, indicating low damping and strong resonance between GFL2 and the grid. The lowest phase margin is at 30~Hz, confirming the strongest coupling-induced resonance. To validate this prediction, a 1\% voltage step is applied to the Thevenin equivalent. Fig.~\ref{fig:9bustimeresponse} shows the time-domain response at buses 4 and 6, highlighting transient oscillations.

Finally, Table~\ref{tab:passive9} summarizes the passivity analysis. GFL2 exhibits non-passive behavior within the 8–30~Hz range, whereas the grid remains largely passive except at 8~Hz. These findings enable system designers to retune control parameters and mitigate high-risk scenarios. Additional tests with varying PLL gains for GFL2 indicate that increasing \(k_i\) shifts the dominant resonance toward higher frequencies, while decreasing \(k_i\) moves it to lower frequencies.

\begin{table}[t!]
\centering
\caption{Summary of passivity results.}
\label{tab:passive9}
\begin{tabular}{|c|c|c|}
\hline
\textbf{System} & \textbf{Frequency Range} & \textbf{Status} \\ \hline
System 1 (GFL2) & 1.0–34 Hz & Non-passive \\ \hline
System 2 (Grid) & 6.0–8.0 Hz & Non-passive \\ \hline
\end{tabular}
\vspace{-5mm}
\end{table}

\vspace{-5mm}

\section{Conclusions}\label{sec:conclusions}
This paper presents and promotes SIaD-Tool, a comprehensive, open-source frequency-domain (FD) tool for stability and interaction assessment in modern power systems. The tool enables scanning across multiple reference frames (\( abc \), \( dq0 \), and \( 0pn \)) using a novel perturbation scheme. It integrates both series voltage and parallel current perturbation strategies, and supports various excitation signals. By allowing direct scanning in the desired reference frame using a steady-state ideal source and decoupling technique, it eliminates the need for complex post-processing to address mirror or coupling frequency artifacts, filling a critical gap in existing methodologies.

The tool has been validated through a wide range of case studies, including converter models (GFL and GFM), an offshore wind power plant, and the IEEE 9-bus system. These applications demonstrate the tool ability to identify stability boundaries and critical interaction scenarios, providing a comprehensive assessment of system dynamics. SIaD-Tool also enables precise characterization and broader analysis of black-box systems, overcoming proprietary limitations and facilitating broader system-level analysis.

The tool is freely available in the GitHub repository \cite{SIaD_tool}, with automated versions for MATLAB/Simulink and Python/PSCAD platforms, promoting accessibility and reproducibility across academia, industry, and grid operators.

\vspace{-4mm}

\section{Acknowledgment}\label{sec:Acknowledgment}
This work has received funding from the ADOreD project of the European Union’s Horizon Europe Research and Innovation program under the Marie Skłodowska-Curie grant agreement No. 101073554.
\vspace{-3.5mm}
\bibliographystyle{IEEEtran}
\bibliography{ref}

\vfill

\newpage

 




\vfill

\end{document}